\documentclass[twocolumn,superscriptaddress,floatfix,preprintnumbers, amsmath, amssymb]{revtex4-2}

\usepackage[dvipsnames]{xcolor}
\usepackage{graphicx}
\usepackage{dcolumn}
\usepackage{bm}
\usepackage{amsfonts}
\usepackage{soul}
\usepackage[justification=justified, font=small]{caption}
\usepackage[justification=justified, font=scriptsize]{subcaption}
\usepackage{comment}
\usepackage{multirow}
\usepackage{braket}

\usepackage{tabularx}
\begin{document}


\title{Error Suppression for Arbitrary-Size Black Box Quantum Operations}

\author{Gideon Lee}
\affiliation{Pritzker School of Molecular Engineering, The University of Chicago, Chicago, Illinois 60637, USA}

\author{Connor T. Hann}
\email{This work was done prior to CTH joining the AWS Center for Quantum Computing.}
\affiliation{Department of Applied Physics, Yale University, New Haven, CT 06520}
\affiliation{Department of Physics, Yale University, New Haven, Connecticut 06511, USA}
\affiliation{Yale Quantum Institute, New Haven, Connecticut 06520, USA}
\affiliation{AWS Center for Quantum Computing, Pasadena, California 91125, USA}
\affiliation{IQIM, California Institute of Technology, Pasadena, CA 91125, USA}

\author{Shruti Puri}
\affiliation{Department of Applied Physics, Yale University, New Haven, CT 06520}

\author{S.M. Girvin}
\affiliation{Department of Physics, Yale University, New Haven, Connecticut 06511, USA}
\affiliation{Yale Quantum Institute, New Haven, Connecticut 06520, USA}

\author{Liang Jiang}
\affiliation{Pritzker School of Molecular Engineering, The University of Chicago, Chicago, Illinois 60637, USA}
\affiliation{AWS Center for Quantum Computing, Pasadena, California 91125, USA}

\date{\today}

\begin{abstract}
Efficient suppression of errors without full error correction is crucial for applications with NISQ devices. Error mitigation allows us to suppress errors in extracting expectation values without the need for any error correction code, but its applications are limited to estimating expectation values, and cannot provide us with high-fidelity quantum operations acting on arbitrary quantum states. To address this challenge, we propose to use error filtration (EF) for gate-based quantum computation, as a practical error suppression scheme without resorting to full quantum error correction. The result is a general-purpose error suppression protocol where the resources required to suppress errors scale independently of the size of the quantum operation, and does not require any logical encoding of the operation.  The protocol provides error suppression whenever an \textit{error hierarchy} is respected -- that is, when the ancilliary cSWAP operations are less noisy than the operation to be corrected. We further analyze the application of EF to quantum random access memory, where EF offers hardware-efficient error suppression.
\end{abstract}

\maketitle

\textit{Introduction -- } One major obstacle to performing meaningful computation on quantum devices is the presence of noise. Canonically, we expect that the theory of fault-tolerant quantum error correction (FTQEC) codes will enable us to scale quantum computers once we have enough qubits and physical error rates fall below a particular threshold \cite{calderbank_good_1996, aharonov_fault-tolerant_1997, gottesman_introduction_2009}.  However, being in the NISQ or early fault-tolerance era means that we are limited in the number and base quality of qubits available \cite{preskill_quantum_2018}, which prevents us from performing full fault-tolerant quantum computing. Recent work related to suppressing errors on NISQ devices has focused on error mitigation \cite{error_mitigation_review} -- for instance: zero noise extrapolation \cite{li_efficient_2017, temme_error_2017}, quasiprobability decomposition and probabilistic error cancellation \cite{temme_error_2017, Brayvi_EM_ECS_paper, Lostaglio_2021_PRL_QROMagic, Suzuki_2022_arxiv_T_mitig, ibm_PEC_2022}, learning-based methods such as Clifford data regression \cite{strikis_learning-based_2021, czarnik_error_2021} deep learning noise prediction \cite{zlokapa_deep_2020}, and virtual distillation \cite{koczor_exponential_2021, huggins_virtual_2021, czarnik_qubit-efficient_2021}. Such methods allow the user to suppress errors in \textit{extracting expectation values} with minimal hardware overhead. The success of such methods in the near-term motivates the desire to \textit{suppress errors in quantum gates} beyond expectation values. 

One approach to achieve more robust quantum gates is to use error detection techniques. In the near-term, one may not want to use the full formalism of QEC. One promising alternative approach to detect errors without full QEC is Error Filtration (EF), which was first introduced as a means to stabilize quantum communication \cite{gisin_error_2005}. EF does not seek to mitigate errors in expectation values but rather to protect quantum information during noisy communication. In essence, EF multiplexes a single message, and then attempts to detect and discard the parts of this message in which errors have occurred. Up to post-selection, one is able to communicate a message  over multiple similarly noisy channels with lower error rates than a single noisy channel. Given a single-channel error that goes as $\varepsilon$, EF is able to suppress errors in the fidelity of the communicated message to $\varepsilon / T$, where $T$ is the number of channels in the multiplexing which corresponds to the  effective dimension of the ancilla Hilbert space. Due to its ease-of-implementation, a successful proof-of-principle experiment was quickly carried out \cite{lamoureux2005experimental}. Recent interest in EF has seen a revival in the context of a more general class of schemes communicating over a quantum superposition of trajectories. Such schemes boast a range of exotic and remarkable results, such as perfect quantum communication over zero-capacity channels \cite{chiribella_quantum_2019, chiribella2021indefinite, kristjansson_witnessing_2021} . Separately, \cite{Vijayan2020robustwstate} also formalized aspects of EF and derived explicit EF fidelities for loss and dephasing channels. However, until now, EF has mostly been studied in the context of suppressing errors in communication, which is restricted to identity operations \footnote{Similar to our work, \cite{Vijayan2020robustwstate} proposed encoding single-qubit unitaries in a Fourier basis to similarly suppress errors. However, in our work, we show that in the gate-based context no such encoding is required, and that the resultant error suppression extends to any number of qubits.}. Also relevant shortly is the development of biased noise qubits and bias-preserving gates which suffer from an exponentially smaller likelihood of bit flips than phase flips \cite{rep_cats, puri2020bias, amazon_cats}. In QEC, the use of biased-noise architectures gives rise to much higher fault-tolerant thresholds for QEC codes \cite{Darmawan_2021, Tuckett_2018}. In the context of this work, they boost the performance of a protocol that might otherwise be infeasible in the near future.


In this Letter we extend EF to the context of gate-based quantum computation (gate-based EF), and show that this provides a low-overhead means to suppress errors for a large class of quantum operations. The result is a general-purpose error detection protocol where  \textit{the resources required to suppress errors scale independently of the size of the quantum operation}. This has the appealing consequence of allowing us to leverage any small number of additional qubits  available in a noisy device to suppress errors in large, complex quantum operations. To establish these results, we provide a general quantum circuit design (Fig. \ref{fig:genM_diag}) that can filter out errors by employing the noisy operations as black boxes. We stress that EF deals with quantum operations, and provides a means to move beyond suppressing expectation values without the full formalism of QEC.

\bigskip

\begin{figure}
    \centering
    \includegraphics[width=\linewidth]{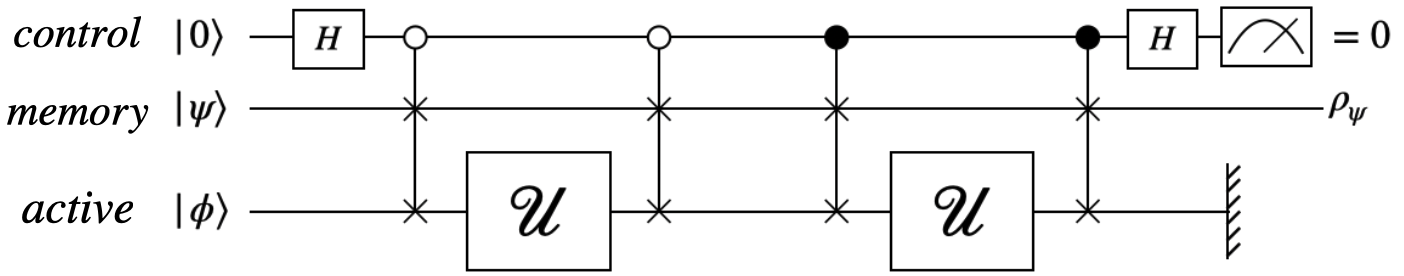}
    \caption{Gate-based EF with a single control qubit. This circuit makes two calls to an apparatus that implements the noisy process $\mathcal{U}$ as follows: (1) Prepare the \textit{control register} in the equal superposition state $|+\rangle = H |0\rangle$, the intended input state $|\psi\rangle$ in the \textit{memory register}, and any easy-to-prepare state $|\phi\rangle$ in the \textit{active register}. (2) Conditioned on the state of the control register being in $|0\rangle$ ($|1\rangle$), the circuit first applies $\mathcal{U}$ to $|\psi\rangle$ ($|\phi\rangle$), then to $|\phi\rangle$ ($|\psi\rangle$). (3) After a final hadamard, the circuit post-selects the result conditioned on obtaining the outcome $|0\rangle$. The active register is discarded. This results in the suppression of the infidelity of the apparatus by half (see Eq.~(\ref{eqn:basecaseinfid})). }
    \label{fig:minEF_diag}
\end{figure}

\textit{Gate-based EF with one control qubit -- } Suppose we are given a black box that imperfectly carries out some ideal unitary $U$. We model this as a CPTP map $\mathcal{U}$ comprising Kraus operators $K_i, i = 0, 1, ..., R$ \cite{nielsen_chuang_2019}, where we define $K_0$ to be the Kraus operator whose normalized action most resembles $U$. We assume that queries to this black box are already fairly close to the ideal unitary $U$.  We can formalize this notion by assuming that $|| K_0 - U || \leq \varepsilon \ll 1$. Then we can write $K_0 = U - \varepsilon \xi$ for some suitably normalized operator $\xi$ with $\| \xi \| = 1$. The infidelity goes as $(1 - F)_0 \equiv \langle U \psi | \mathcal{U} (\psi) | U \psi\rangle \sim O(\varepsilon)$ for non-unitary errors, while $(1 - F)_0 \sim O(\varepsilon^2)$ for unitary errors \cite{SM}. Without any further information, or prying the box apart and subjecting every qubit to QEC, how can we suppress the $O(\varepsilon)$ non-unitary error? We argue that one effective way to do so, leveraging a small number of high quality biased-noise qubits, is gate-based EF.

We begin by introducing gate-based EF with only one control qubit. This provides a gentle introduction to gate-based EF in its simplest possible incarnation and demonstrates that \textit{with minimal overhead, gate-based EF is able to help us achieve non-trivial error suppression in the near-term}. One can think about multiplexing in the original EF protocol as quantum communication over a superposition of trajectories \cite{chiribella_quantum_2019}. Inspired by this structure, we want to create a superposition of $T$ queries to $\mathcal{U}$. Fig.~\ref{fig:minEF_diag} depicts the minimal implementation of gate-based EF with $T = 2$. To create a superposition of calls to the black box, we need three ingredients. The first is entanglement with a single qubit \emph{control register} that maintains the superposition between calls to $\mathcal{U}$, the second is a \emph{memory register}, initialized with the desired input state $| \psi \rangle$, to store the results of these calls, and the third is an \emph{active register}, initialised in some (for now) arbitrary state $| \phi \rangle$, for null calls to $\mathcal{U}$, which is discarded at the end of the protocol. We prepare the control register in the equal superposition state $| + \rangle$. Conditioned first on the control register being $|0 \rangle$ and then $|1\rangle$, we query $\mathcal{U}$ twice with the input state $| \psi \rangle$. Subsequently, we take a measurement on the control register and post-select on obtaining $| + \rangle$. Finally, we trace out the active register, and are left with the final state $\rho_1$. 

To evaluate the scheme, we use the infidelity
\begin{equation}
    (1 - F)_1 \equiv  1 - \frac{\langle U \psi | \rho_{1} | U \psi \rangle}{\mbox{Tr} \rho_{1}},
    \label{eqn:infid1}
\end{equation}
where $|U \psi \rangle = U | \psi \rangle$ is the ideal state we are trying to achieve. While the above infidelity appears state-dependent, $(1 - F)_1$ as a function of $(1 - F)_0$ turns out to be independent of $| \psi \rangle, | \phi \rangle$. 

First, suppose all the errors in the circuit come from the black box implementation of $\mathcal{U}$. Due to post-selection, $\rho_1$ is not normalized, with $P_1^{(S)} \equiv  \mbox{Tr} \rho_1$ giving the success probability. 
A straightforward calculation gives the unnormalized state 
\begin{equation}
\rho_1 = \frac{1}{2} \mathcal{U} ( | \psi \rangle  \langle \psi |) + \frac{1}{2} \sum_{i = 0, j = 0}^{R}  K_i  | \psi \rangle \langle \psi | K_j^{\dagger} {\rm Tr} \left( \rho_{\phi} K_i^{\dagger} K_j \right),
\label{eqn:M2rhoraw}
\end{equation}
and success probability,
\begin{equation}
  P_s^{(1)} = \frac{1}{2} + \frac{1}{2} \sum_{i = 0, j = 0}^{R}   \langle \psi | K_j^{\dagger} K_i | \psi \rangle {\rm Tr} \left( \rho_{\phi} K_i^{\dagger} K_j \right).
\end{equation}

Recall our assumption that $K_0 = U - \varepsilon \xi$, which implies that the terms in the sum where $i, j \neq 0$ are $O( \varepsilon^2  )$. As such, when computing the fidelity we only have to keep terms where either $i, j = 0$. For clarity of presentation, we now assume the simplest possible model fulfilling our assumptions, with $K_0 = \sqrt{1 - \varepsilon} U, K_1 = \sqrt{\varepsilon} V$, and $V \neq U$ some arbitrary erroneous unitary. We deal with the general case in \cite{SM}, for which the main result below (Eq.~(\ref{eqn:basecaseinfid})) has exactly the same form. Specializing to this error model, the explicit probability of success is close to $1$: 
\begin{equation}\label{eqn:basecaseprob}
    \begin{aligned}
  &1 - P_1^{(S)} \\
  &\simeq \varepsilon \left(  1 - {\rm Re} \left\{ \langle \psi | U^{\dag} V | \psi \rangle {\rm Tr} ( \rho_{\phi} V^{\dag} U )\right\}\right) 
    \leq 2 \varepsilon
\end{aligned}
\end{equation}
where we have discarded $O(\varepsilon^2)$ terms, and the final inequality comes from unitarity of $U, V$. The inner product with the ideal state is,
\begin{equation}\label{eqn:basecaseoverlap}
\begin{aligned}
  &\langle \psi | U^{\dag} \rho_1 U | \psi \rangle \\
  &\simeq \frac{1}{2} F_0 + \frac{1}{2} 
  - \varepsilon \left(  1 - {\rm Re} \left\{ \langle \psi | U^{\dag} V | \psi \rangle {\rm Tr} ( \rho_{\phi} V^{\dag} U )\right\}\right).
\end{aligned}    
\end{equation}

Inserting Eqs.~(\ref{eqn:basecaseprob}, \ref{eqn:basecaseoverlap}) into (Eq.~\ref{eqn:infid1}), the infidelity is readily obtained. Up to $O(\varepsilon)$, the $V, \phi$ dependent terms cancel out, we have
\begin{equation}\label{eqn:basecaseinfid}
    \begin{aligned}
        (1 - F)_1
  &= \frac{1}{2} (1 - F)_0 + O ( \varepsilon^2),
    \end{aligned}
\end{equation}
i.e. the infidelity is halved with a single ancilla qubit. Eq.~(\ref{eqn:basecaseinfid}) is our first key result, and demonstrates the ability of gate-based EF to suppress errors with minimal overhead comprising of a single control qubit and an additional memory register. We highlight that the scaling of $(1 - F)_1$ with $(1 - F)_0$ is independent of the erroneous unitary $V$ or the active register state $|\phi\rangle$. This latter point suggests that we may initialize the active register in any state, eg. a thermal state, that is easiest to prepare in the lab. Finally, note that as long as $F_0 > 1/2$, gate-based EF can still provide error suppression even if $\varepsilon \sim O(1)$ \cite{SM}. 

\bigskip

\textit{The effects of ancilla noise -- }  In the near-term, we will not have perfectly noiseless ancillas. What happens to the results Eq.~(\ref{eqn:basecaseprob}, \ref{eqn:basecaseinfid})? We expect the primary mechanism of introducing additional errors to be the cSWAP gate. One mitigating factor is the recent discovery of biased-noise cat codes allow us to construct a bias-preserving Toffoli gate \cite{rep_cats, puri2020bias, amazon_cats}, from which we can construct controlled-SWAP (cSWAP) operations in a bias-preserving manner. Recent experimental work \cite{reglade_2023_AandB, Grimm_Frattini_Puri_Mundhada_Touzard_Mirrahimi_Girvin_Shankar_Devoret_2020} has demonstrated biases of up to $10^{9}$ can be achieved with cat qubits, and theoretical work \cite{Xu_Zheng_Wang_Zoller_Clerk_Jiang_2023} demonstrates that such extreme biases can be achieved with modest excitation numbers and low dominant error rates. We thus focus on ancilliary phase flip errors, the effects of which are two-fold. First, a phase flip error on the control register commutes with the cSWAP operation. Hence it can be propagated to the end of the circuit, where it is detected by the measurement -- i.e. phase flip errors decrease the success probability of the scheme without affecting the fidelity. Assuming this event occurs with some probability $p_z$ over the course of the circuit, we can modify Eq.~(\ref{eqn:basecaseprob}) by substituting $P_1^{(S)} \rightarrow P_1^{(S)} - p_z$. 

On the other hand, errors on the memory register are more problematic, since they interact with the input state $|\psi\rangle$ in an uncontrolled way. Let $p_m$ be the effective probability of a memory register error during the protocol. Since most near term devices are dominated by gate errors \cite{Google_paper_2014}, we can assume that these errors come from the cSWAP, so that $p_m \sim p_z$. The worst case assumption that this error completely ruins the query and cannot be detected by post-selection modifies Eq.~(\ref{eqn:basecaseinfid}) by $(1-F)_1 \rightarrow (1-F)_1 + p_m$. This reveals the ideal operation of gate-based EF to require an \textit{error hierarchy}. Eq.~(\ref{eqn:basecaseinfid}) holds when $p_m \ll \varepsilon \ll 1$. However, as long as $p_m < (1 - F)_0 \sim \varepsilon$, gate-based EF will still suppress errors if the cSWAP operations are less noisy than $\mathcal{U}$ \cite{SM}. 

\begin{figure*}[htpb]
    \centering
    \includegraphics[width=0.9\linewidth]{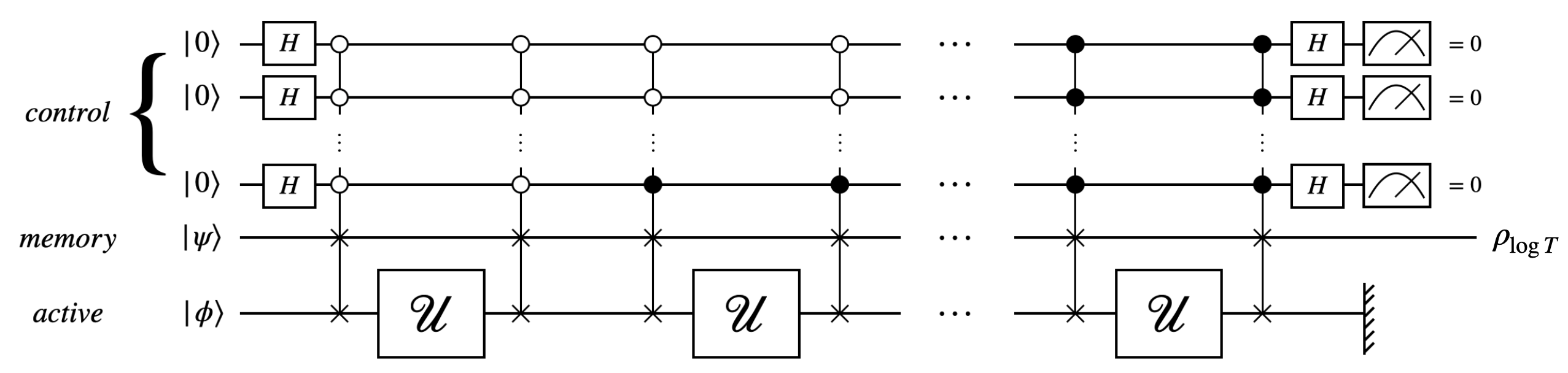}
    \caption{General circuit for gate-based EF that makes some number $T$ calls to an apparatus that implements the noisy process $\mathcal{U}$. This generalizes the circuit in Fig.~\ref{fig:minEF_diag} by allowing for $\log T$ control qubits, thus making $T$ calls to $\mathcal{U}$ conditioned on each branch $| i \rangle$ of the control register in the computational basis (i.e. $|0 ... 00 \rangle, | 0 ... 01 \rangle, \ldots$).}
    \label{fig:genM_diag}
\end{figure*}

\bigskip

\textit{Gate-based EF with $T$ control qubits -- } Having introduced the base case of gate-based EF applicable to near-term devices, we now generalize the notion of gate-based EF to a situation with $\log T$ control qubits \footnote{ Note that $\log$ is taken to base $2$ in this text, such that $\log T$ refers to the number of control qubits utilized.}. While the $\log T = 1$ case examined earlier is immediately applicable, we envision the general case to be useful when better ancilla qubits become available. To motivate this, one can imagine having a small number of high quality or error-corrected qubits, which do not suffice for full QEC of the apparatus $\mathcal{U}$. However, applying these qubits as ancillas in gate-based EF, one can still achieve considerable error suppression for $\mathcal{U}$.

The generalization is depicted in Fig.~\ref{fig:genM_diag}. The two modifications to the base case circuit are appending of additional qubits to the control register, and the usage of many-cSWAPs. In this case, we have $T$ applications of $\mathcal{U}$ to $| \psi \rangle$ conditioned on the control qubits being in $|00...00\rangle, |00...01\rangle, ..., |11 ... 11\rangle$.

When the ancillas are much less noisy than the apparatus, Eq.~(\ref{eqn:basecaseinfid}) generalizes to,
\begin{equation}
\begin{aligned}
(1 - F)_{\log T} 
& \simeq  \frac{1}{T} (1 - F)_0 + O(\varepsilon^2),
\end{aligned}
\label{eqn:infid_scaling}
\end{equation}
 independent of $| \psi \rangle, |\phi \rangle$. This is a key result of our work, and the full derivation is contained in \cite{SM}. 

To understand the scaling of the success probability with $T$, we can make the worst-case assumption that the occurrence of an error $K_{i > 0}$ on any step causes us to reject the output. This yields the lower bound 
\begin{equation}
    P_{\log T}^{(S)} \geq 1 -  T\varepsilon + O(\varepsilon^2)
    \label{eqn:gen_pf_scaling}
\end{equation}
When $T \ll 1/\varepsilon$, the scheme will still work with high probability, and Eq.~(\ref{eqn:gen_pf_scaling}) allows one to trade-off between error suppression and success probability. 

Surprisingly, we can often do better than Eq.~(\ref{eqn:gen_pf_scaling}). Under certain favorable conditions, the success probability can be lower-bounded by a constant,
    \begin{equation}
        \begin{aligned}
        P_{\log T}^{(S)} \geq 1 - 4 \varepsilon + \frac{\varepsilon}{T}.
        \end{aligned}
    \end{equation}
    which approaches a constant $P_{\log T}^{(S)} \rightarrow 1 - 4\varepsilon$ as $T$ increases. These conditions are detailed in \cite{SM}, but we note a particularly relevant case: if the apparatus $\mathcal{U}$ has biased ($Z$) noise, one can achieve this bounded success probability by initializing the active register in the $Z = +1$ eigenstate $|00 \ldots 0 \rangle$.

    Finally, as long as ancilla errors remain smaller than the apparatus errors, one can still find an optimal point for $T$ where the error suppression of gate-based EF is maximum \cite{SM}.

\bigskip

\textit{Application to QRAM -- } To illustrate the practical utility of our scheme, we consider its application to quantum random access memory (QRAM) \cite{giovannetti_architectures_2008, arunachalam2015robustness}. Error-correcting QRAM has an extremely large hardware overhead. The base hardware overhead of QRAM is already tremendous -- in order to prepare a state on $\log N$ address bits, QRAM requires $O(N)$ physical qubits. In data processing applications, relevant values of $N$ could easily reach $N \sim 10^6 - 10^9$ individually encoded qubits. This problem of hardware overhead is compounded by QRAM being a non-Clifford operation, requiring special techniques \cite{universal_FT, Hill_2011_codeswitching} to implement fault-tolerantly \cite{nielsen_chuang_2019}. The detailed analysis of \cite{FT_QRAM} corroborates this intuition by demonstrating that a fault-tolerant surface code implementation of QRAM for a memory of size $N \sim 10^6 - 10^9$ would require some $ 10^{10} - 10^{13}$ physical qubits.

In contrast to QEC, the resource overhead of gate-based EF scales independently of the size of the desired quantum operation. Additionally, QRAM satisfies the two conditions for the optimal application of gate-based EF. First, the error hierarchy is enforced since $N$ is generally very large, ensuring many more errors occur in the apparatus than the ancillas. Second, one can implement QRAM with biased noise, satisfying the conditions for an upper-bounded failure probability \cite{SM}. This ensures that it remains feasible to embed a QRAM with gate-based EF into a quantum algorithm as an oracle. Thus, gate-based EF can suppress QRAM errors in a hardware-efficient manner. For a more complete review of QRAM and the numerical techniques used to simulate it, see \cite{hann_resilience_2021}.

\begin{figure}
    \centering
    \includegraphics[width=\linewidth]{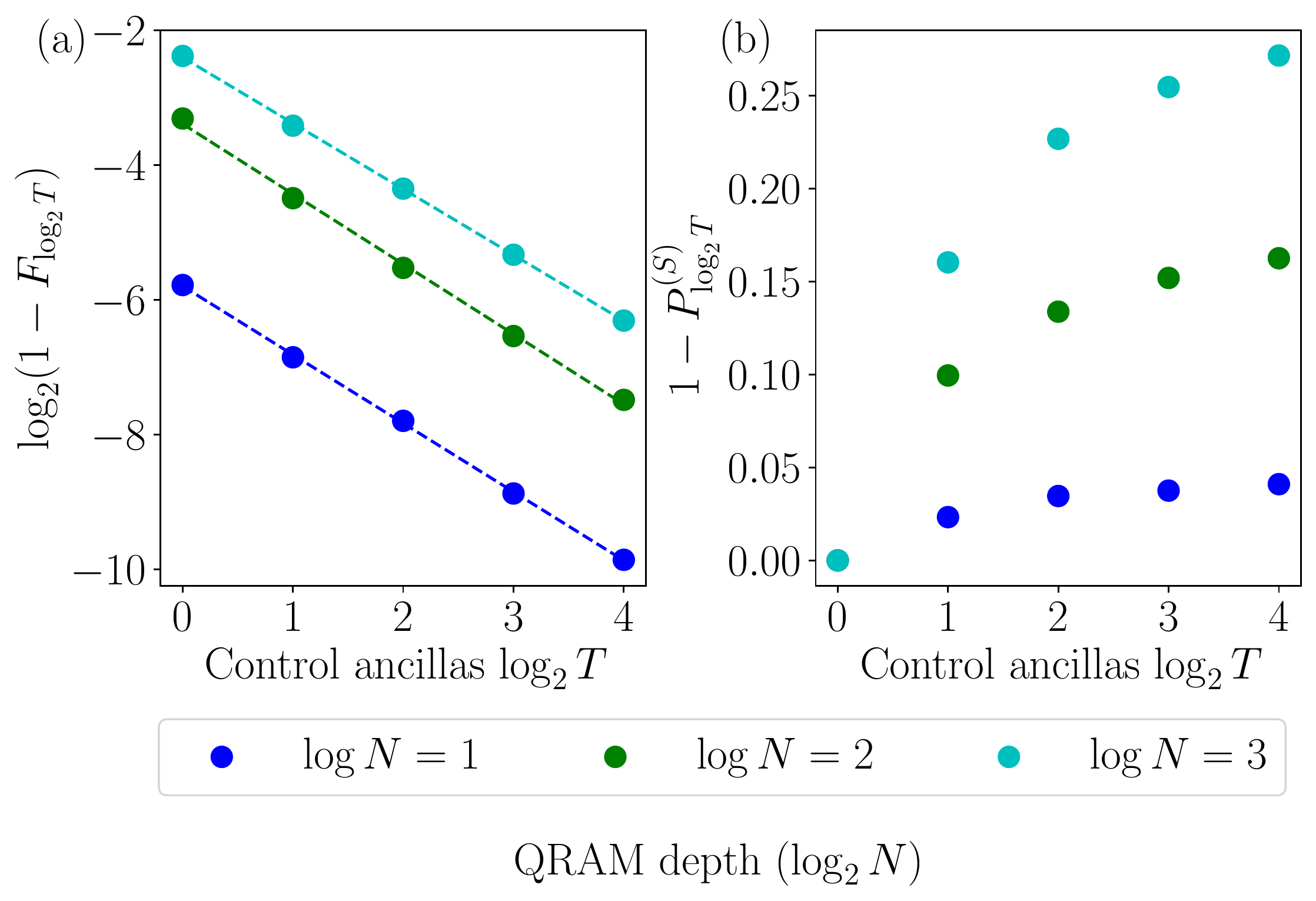}
    \caption{Gate-based EF applied to QRAM subject to depolarizing errors. (a): Plot of $\log(1 - F_{\log T})$ as a function of $\log T$, where $F_{\log T}$ denotes QRAM query fidelity obtained after gate-based EF with $\log T$ control qubits. The dashed lines indicate linear fits. This demonstrates good agreement with the expected $1/T$ suppression for low apparatus error, with the deviation of the estimated slope from $-1$ explained by the $O(\varepsilon^2)$ terms for higher apparatus error rates. (b): Plot of failure probabilities $1 - P_{\log T}^{(S)}$ as a function of $\log T$. The simulated failure probability for depth $1, 2$ QRAMs quickly plateau. All failure probability plots show sub-linear scaling, as expected.}
    \label{fig:QRAM_scaling}
\end{figure}

To showcase our scheme's hardware efficiency, we numerically simulate its application to QRAM circuits comprising up to $2^n = 8$ qubits with up to $T = 4$ ancilla qubits in Fig. \ref{fig:QRAM_scaling} without ancilla errors. Absent ancilla errors, we see an excellent agreement with the $1/T$ scaling. With ancilla errors \cite{SM}, we find that there is some $T$ for which gate-based EF is optimal. We simulate the physical qubits of the QRAM with a $0.01$ depolarizing error rate per time step, which gives a base infidelity that goes as $O(0.01 (\log N)^2)$ \cite{hann_resilience_2021}. In this regime, we find that our scheme can reduce the query infidelity by over an order of magnitude using only $4 + \log N$ additional qubits. For QRAM, we have shown that one can begin to suppress errors, albeit to a smaller extent than QEC, with only $\log T + \log N$ additional qubits. Our scheme thus provides a compelling alternative to QEC in the NISQ and pre-FTQEC eras.

\bigskip

\textit{Hardware efficiency -- } In this section, we discuss hardware overhead. First, to obtain a $1/2$ suppression, we must append a memory register and an ancilla qubit. The size of the memory register depends on the size of the \textit{input} to $\mathcal{U}$. We stress that this is not the same as the size of $\mathcal{U}$. Besides QRAM, many useful quantum operations, are only possible with an (often exponentially) large number of ancillas --  fast state preparation on $n$ qubits in $O(n)$ time requires $O(2^n/n)$ ancillas \cite{gui2023spacetimeefficient, Yuan_2023}. Since each qubit of the memory register must be swapped with the active register, this linearly increases the number of cSWAPs. By swapping each qubit in sequentially, we can avoid adding to qubit overhead. To go from a $1/2$ to a $1/T$ suppression, one appends $\log T$ qubits \textit{to the control register only} -- additional error suppression has a hardware overhead that scales largely independently of $\mathcal{U}$.

A further hardware benefit is that error suppression from gate-based EF is largely agnostic to $\mathcal{U}$. This sidesteps any complications of having to construct \textit{logical} versions of $\mathcal{U}$, since the \textit{physical} implementation will do. In particular, error suppression via gate-based EF is \textit{independent of either the
complexity of operating the black box or knowledge of
the ideal unitary to be carried out.} QRAM aside, state preparation also requires $\Omega(\sqrt{2^n \log n})$ $T$ gates \cite{low2018trading}. This non-Cliffordness compounds the hardware overhead of QEC, but does not affect gate-based EF.

\bigskip
\textit{Discussion -- } In this work, we proposed and analyzed the circuit implementation of error filtration on a qubit quantum computer in two levels. In \cite{SM}, we show that gate-based EF is isomorphic to the original error filtration setup in the noiseless case. However, our analysis considers more general error models than the loss and dephasing models considered in \cite{gisin_error_2005, Vijayan2020robustwstate}. We emphasize the hardware-efficiency of gate-based EF, which allows one to use a small number of qubits with resources scaling independently of the quantum operation considered to suppress errors up to a quadratic error floor. We further emphasize that in contrast to usual error mitigation schemes, gate-based EF suppresses errors in quantum gates, which extends the reach of error mitigation to problems involving state preparation and sampling. As a comparison, one class of schemes that performs error suppression for unitary operations is the extended flag gadget scheme \cite{Debroy_2020_extended_flag, PCS_paper_pub}. However, such schemes have limited capability in suppressing errors in non-Clifford circuits. Conversely, our scheme is entirely agnostic to the structure of the desired quantum process -- in particular, QRAM is non-Clifford. Gate-based EF extends the reach of low-overhead error suppression methods, which is an important step towards bridging the NISQ and fault-tolerance eras. 

While we have characterized gate-based EF by reference to NISQ-era error mitigation schemes, we note that many-controlled SWAP gates may not be easy to implement on NISQ devices. As such, one might regard gate-based EF as a scheme most suited for a post-NISQ era but before achieving FTQEC. However, this does not completely rule out the application of gate-based EF during the present NISQ era in the $T = 2$ case. In the microwave regime, a high fidelity $(>0.95)$ cSWAP recently been reported in \cite{chapman_cSWAP}. Alternatively, optical implementations may also be suitable for gate-based EF (see eg. \cite{chen_scalable_2021, Hong_PRA_2012_robust_QRAM, ono2017implementation} as well as the SM \cite{SM} for further discussion).

\begin{acknowledgments}
We thank Senrui Chen, Ng Hui Khoon, Qian Xu, and Kaiwen Gui for helpful discussions. We acknowledge support from the ARO (W911NF-18-1-0020, W911NF-18-1-0212), ARO MURI (W911NF-16-1-0349, W911NF-21-1-0325), AFOSR MURI (FA9550-19-1-0399, FA9550-21-1-0209), AFRL (FA8649-21-P-0781), DoE Q-NEXT, NSF (OMA-1936118, ERC-1941583, OMA-2137642), NTT Research, and the Packard Foundation (2020-71479).
\end{acknowledgments}


\bibliography{apssamp}


\clearpage


\widetext
\begin{center}
\textbf{\large Supplemental Material: "Error Suppression for Arbitrary-Size Black Box Quantum Operations"}
\end{center}
\setcounter{equation}{0}
\setcounter{figure}{0}
\setcounter{table}{0}
\setcounter{page}{1}
\makeatletter
\renewcommand{\theequation}{S\arabic{equation}}
\renewcommand{\thefigure}{S\arabic{figure}}
\renewcommand{\bibnumfmt}[1]{[S#1]}
\renewcommand{\citenumfont}[1]{S#1}

\section{Noise Model}

We assume we are given an apparatus that is meant to carry out the unitary $U$ but instead carries out a CPTP channel $\mathcal{U}( \cdot )$, with $R + 1$ Kraus operators $\left\{ K_0, K_1, ..., K_R \right\}$, where $K_0$ is the Kraus operator closest to $U$. Now, in our setup we append ancillas to this apparatus. Let $I_A$ be the collective identity operation on all parts of the system that aren't part of the apparatus. Then, assuming the ancillas are noiseless, a reasonable way to extend the noise model to the rest of the system is to use Kraus operators
\begin{equation}
K^{(s)}_i = I_A \otimes K_i, i = 0, 1, 2, ..., R.
\end{equation}
This assumption is justified when the cross-talk between the apparatus and the rest of the system is small. Subsequently, we will suppress the $I_A$ and $(S)$ superscript, writing $K_i$ to be understood in the above sense.

\section{Small parameter case}

\begin{figure*}[htpb]
    \centering
    \includegraphics[width=0.9\linewidth]{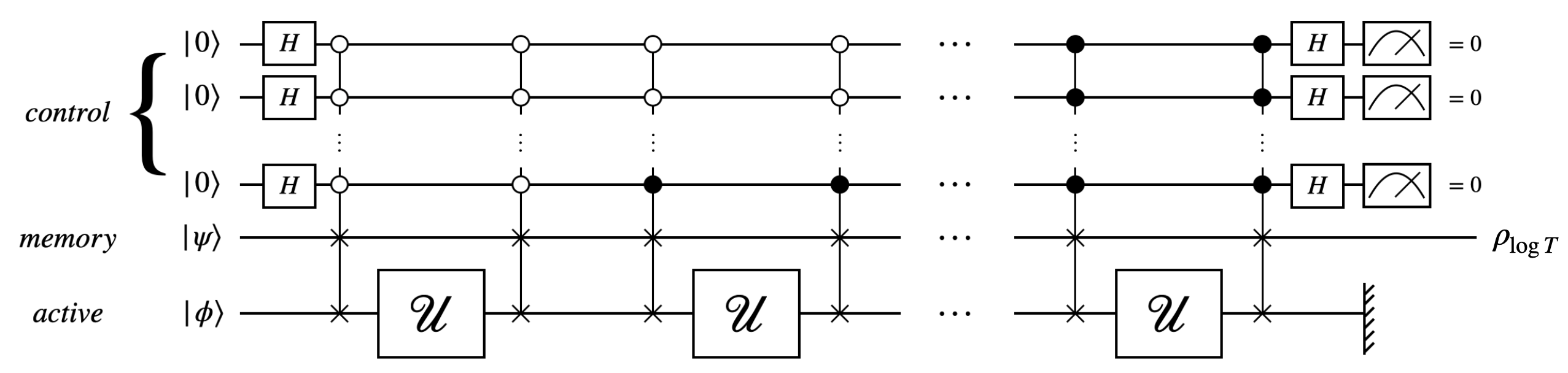}
    \caption{General circuit for gate-based EF that makes some number $T$ calls to an apparatus that implements the noisy process $\mathcal{U}$ as follows: (1) Prepare the intended input state $| \psi \rangle$ in one register, and the null state $ | \phi \rangle$ in another. We refer to the former as the \textit{memory register} and the latter as the \textit{active register}. (2) Prepare $\log T$ ancillas in the \textit{control register} in the equal superposition state $| + + ... + \rangle$. (3) For each branch $| i \rangle$ in the computational basis (i.e. $|0 ... 00 \rangle, | 0 ... 01 \rangle$, and so forth), swap the state  $| \psi \rangle$ into the active register. We then run the active register through our apparatus, given by noisy process $\mathcal{E}$, before swapping the registers back for the same branch. Do this for every branch of the superposition. (4) Finally, perform $\log T$ parallel measurements on the control register in the $X$ basis, and post-select for every measurement being $+1$. This corresponds to projecting the control register back onto the equal superposition state $| + + ... + \rangle$.}
    \label{fig:genM_diag_SM}
\end{figure*}

We begin by deriving the scaling behaviour of gate-based EF in the primary case of interest -- that is, we are given an apparatus that is already pretty good, and everything else in the circuit is noiseless. Formally, we say that there exists a Kraus operator $K_0$ that is $\varepsilon$-close to the ideal unitary $U$, such that we can write $K_0 = U - \varepsilon \xi$ for some small $\varepsilon$ and some suitably normalized operator $\xi$ with $\| \xi \| = 1$. Given such a form for $K_0$, it follows that $K_i \sim O( \sqrt{\varepsilon})$ for $i \neq 0$. We will also assume that $1 - F_0 \sim O(\varepsilon)$, where $F_0$ is the native fidelity of the apparatus prior to error fitration. Note that this assumption is not compatible with $K_0 = V$ where $V$ is some coherent unitary error, as we will see in Sec.~\ref{SM_sec:coherent_error}, where we consider those cases separately.

\subsection{$T = 2$ case}

We will first work out the $T = 2, \log T = 1$ case. We index all the relevant density matrices and fidelities with $\log T$. The unnormalized output of Fig.~\ref{fig:genM_diag_SM} when $\log T = 1$ is the density matrix
\begin{equation}
    \tilde{\rho}^{(1)} = \frac{1}{2} \mathcal{U} ( | \psi \rangle  \langle \psi |) + \frac{1}{2} \sum_{i = 0, j = 0}^{R}  K_i  | \psi \rangle \langle \psi | K_j^{\dagger} {\rm Tr} \left( \rho_{\phi} K_i^{\dagger} K_j \right),
\label{eqn:M2rhoraw}
\end{equation}
where we note that $F_0 = \langle \psi | U^{\dag} \mathcal{U} ( | \psi \rangle  \langle \psi |)  U | \psi \rangle$. The conclusions of this section are independent of the state $ | \phi \rangle$, so we can choose any $ | \phi \rangle$.

The probability of success is given by the trace of this density matrix,
\begin{equation}
  P_s^{(1)} = \frac{1}{2} + \frac{1}{2} \sum_{i = 0, j = 0}^{R}   \langle \psi | K_j^{\dagger} K_i | \psi \rangle {\rm Tr} \left( \rho_{\phi} K_i^{\dagger} K_j \right).
\end{equation}

Recall our assumption that $K_0 = U - \varepsilon \xi$. By the arguments in the previous section, the terms in the sum where $i, j \neq 0$ are $O( \varepsilon^2  )$. As such, when computing the fidelity we only have to keep terms where either $i, j = 0$. We first show that the probability of success is close to $1$: 
\begin{align*}
  P_s^{(1)} = {\rm Tr} \tilde{\rho}^{(1)}
  &=  \frac{1}{2} + \frac{1}{2} \langle \psi | K_0^{\dag} K_0 | \psi \rangle  {\rm Tr} (\rho_{\phi} K_0^{\dag} K_0) + \frac{1}{2} \sum_{i \neq 0} \left( \langle \psi | K_0^{\dag} K_i | \psi \rangle {\rm Tr} (\rho_{\phi} K_i^{\dag} K_0) + h.c. \right) + O( \varepsilon^2)\\
  &= 1
    - \frac{1}{2} \varepsilon \left(  \langle \psi | \xi^{\dag} U | \psi \rangle +  \langle \psi | U^{\dag} \xi | \psi \rangle  +   {\rm Tr} (\rho_{\phi} \xi^{\dag} U) +  {\rm Tr} (\rho_{\phi} U^{\dag} \xi)\right)
    \\ & \quad \quad
    + \frac{1}{2} \sum_{i \neq 0} \left( \langle \psi | U^{\dag} K_i | \psi \rangle {\rm Tr} (\rho_{\phi} K_i^{\dag} U) + h.c. \right) + O( \varepsilon^2)
\end{align*}
which is $\varepsilon$-close to $1$. Similarly, working out the inner product with the ideal state, 
\begin{align*}
  \langle \psi | U^{\dag} \tilde{\rho}^{(2)} U | \psi \rangle
  &= \frac{1}{2} F_0
    + \frac{1}{2}  \langle \psi | U^{\dag} K_0  | \psi \rangle \langle \psi | K_0^{\dagger} U | \psi \rangle {\rm Tr} \left( \rho_{\phi} K_0^{\dagger} K_0 \right)
    \\ & \quad \quad + \frac{1}{2} \sum_{i = 1}^{R}
         \left( \langle \psi | U^{\dag} K_i  | \psi \rangle \langle \psi | K_0^{\dagger} U | \psi \rangle {\rm Tr} \left( \rho_{\phi} K_i^{\dagger} K_0 \right) + \langle \psi | U^{\dag} K_0  | \psi \rangle \langle \psi | K_i^{\dagger} U | \psi \rangle {\rm Tr} \left( \rho_{\phi} K_0^{\dagger} K_i \right)  \right) + O (\varepsilon^2) \\
  &= \frac{1}{2} F_0 + \frac{1}{2} - \frac{1}{2} \varepsilon
    \left( \langle \psi | U^{\dag} \xi  | \psi \rangle  + \langle \psi | \xi^{\dagger} U | \psi \rangle  +  {\rm Tr} \left( \rho_{\phi} \xi^{\dagger} U \right) +  {\rm Tr} \left( \rho_{\phi} U^{\dagger} \xi \right) \right)
  \\ & \quad \quad +
        \frac{1}{2} \sum_{i = 1}^{R}
         \left( \langle \psi | U^{\dag} K_i  | \psi \rangle  {\rm Tr} \left( \rho_{\phi} K_i^{\dagger} U \right) + h.c.  \right) + O (\varepsilon^2)
\end{align*}

Dividing and expanding the denominator in $\varepsilon$, we see that all the remaining $O(\varepsilon)$ terms cancel out, and we are left with
\begin{align}
  (1 - F)_1
  &= 1 - \frac{ \langle \psi | U^{\dag} \tilde{\rho}^{(2)} U | \psi \rangle}{{\rm Tr} \tilde{\rho}^{(2)}} \\
  &= \frac{1}{2} (1 - F)_0 + O ( \varepsilon^2),
\end{align}
i.e. the infidelity is halved with a single ancilla qubit.

\subsection{General $T$ case}

Consider a particular trajectory for a state $| \psi \rangle$ that goes through the circuit with $\log T$ ancilla qubits. At each call of the apparatus it gets a different Kraus operator applied to it, which we write $K_{i_1}, K_{i_2}, \ldots, K_{i_T}$. Whether this operator is applied to the state $| \psi \rangle$ or $| \phi \rangle$ depends on the branch of the ancilla. For the $| t \rangle$ branch of the ancilla, $K_{i_t}$ is applied to $| \psi \rangle$ whereas the rest are applied to $ | \phi \rangle$. We can write this as,
\begin{equation}
| t \rangle | \psi \rangle | \phi \rangle \mapsto | t \rangle K_{i_t} | \psi \rangle K_{i_{T}} K_{i_{T - 1}} \ldots K_{i_{t + 1}} K_{i_{t - 1}} \cdots K_{i_1} | \phi \rangle.
\end{equation}
for  $t = 0, 1, 2, ..., T - 1$, with $|t \rangle$ corresponding to the binary representation of $t$. We can think of $\mathbf{i}$ as a vector with $T$ components $i_j \in \{0, 1, ..., R\}$. The entire vector $\mathbf{i}$ indexes one possible outcome of the circuit.

For ease of notation we will define, 
\begin{equation}
\overline{K_{i_t}} =  K_{i_T} K_{i_{T - 1}} \ldots K_{i_{t + 1}} K_{i_{t - 1}} \cdots K_{i_1}
\end{equation}

To get the full density matrix, we must sum over all possible vectors $\mathbf{i} \in \left\{ 0, 1, ..., R \right\}^{\otimes T}$. This gives the density matrix after the $T$ applications of the black box as,
\begin{equation*}
  \sum_{ \mathbf{i}} \left[ \frac{1}{\sqrt{T}}\sum_{t = 1}^{T} | t \rangle K_{i_t} | \psi \rangle \overline{K}_{i_t} | \phi \rangle \right] \left[ \frac{1}{\sqrt{T}}\sum_{t = 1}^{T} | t \rangle K_{i_t} | \psi \rangle \overline{K}_{i_t} | \phi \rangle \right]^{\dag}
\end{equation*}

After measuring the control register, post-selecting on $|+ \rangle^{\otimes \log T}$ and tracing out the active register,
\begin{align*}
\tilde{\rho}^{(\log T)}
  = \frac{1}{T^2} \sum_{\mathbf{i}} \sum_{t = 1}^{T}  \sum_{q = 1}^{T} K_{i_t} | \psi \rangle \langle \psi | K_{i_q}^{\dag} {\rm Tr} \rho_{\phi} \overline{K}_{i_q}^{\dag} \overline{K}_{i_t}
\end{align*}

First we consider the terms where $t = q$. These give
\begin{align*}
\frac{1}{T^2} \sum_{t = 1}^{T}  \sum_{\mathbf{i}} K_{i_t} | \psi \rangle \langle \psi | K_{i_t}^{\dag} {\rm Tr} \rho_{\phi} \overline{K}_{i_t}^{\dag} \overline{K}_{i_t}
\end{align*}
where the sum over $\mathbf{i}$ can be refactored into 
\begin{align*}
  \sum_{\mathbf{i}} K_{i_t} | \psi \rangle \langle \psi | K_{i_t}^{\dag} {\rm Tr} \rho_{\phi} \overline{K}_{i_t}^{\dag} \overline{K}_{i_t} 
  &= \sum_{j_1 = 0}^R K_{j_1} | \psi \rangle \langle \psi | K_{j_1}^{\dag} \sum_{j_{T} = 0}^R \ldots \sum_{j_2 = 0}^R {\rm Tr} \rho_{\phi} K_{j_{2^T}}^{\dag} \ldots K_{i_2}^{\dag} K_{i_2} \ldots K_{i_{2^T}}  \\
  &=  \sum_{j_1 = 0}^R K_{j_1} | \psi \rangle \langle \psi | K_{j_1}^{\dag} \\
  &= \mathcal{U}(  | \psi \rangle \langle \psi | ),
\end{align*}
where we have used completeness property of the Kraus channel. Hence the density matrix can be rewritten 
\begin{align*}
  \tilde{\rho}^{(\log T)} = \frac{1}{T} \mathcal{U}(  | \psi \rangle \langle \psi | ) 
                    + \frac{1}{T^2} \sum_{\mathbf{i}} \sum_{t = 1}^{T}  \sum_{q \neq t} \left( K_{i_t} | \psi \rangle \langle \psi | K_{i_q}^{\dag} {\rm Tr} \rho_{\phi} \overline{K}_{i_q}^{\dag} \overline{K}_{i_t} \right)
\end{align*}

\subsubsection{General Success Probability}

The success probability is given by 
\begin{equation}
  P_s^{(\log T)} = {\rm Tr} \tilde{\rho}^{(T)}
  = \frac{1}{T} + \frac{1}{T^2} \sum_{\mathbf{i}} \sum_{t = 1}^{T}  \sum_{q \neq t} \left(  \langle \psi | K_{i_q}^{\dag}K_{i_t} | \psi \rangle {\rm Tr} \rho_{\phi} \overline{K}_{i_q}^{\dag} \overline{K}_{i_t}  \right)
\end{equation}

In the general case, we can still get some idea of how the probability of success scales with $T$. Generically we expect the probability to go down with each error and the number of errors to scale with $T$, so we can immediately write down 
\begin{equation}
P_S^{(T)} \sim 1 - O(T) \varepsilon
\end{equation} 

Working through the above expressions more carefully, one finds that the exact scaling at small $\varepsilon$ can be rigorously lower bounded by
  \begin{equation}
P_S^{(T)} \geq 1 - T \varepsilon.
\end{equation}

As long as $T \ll \frac{1}{\varepsilon}$, we will have $P_S^{(T)}$ is $\varepsilon$-close to $1$, we can proceed to calculate the infidelity. In fact, much tighter bounds can be given for $P_S^{(T)}$ under some conditions on the action of the Kraus operators $K_i$ on the state $| \phi \rangle$.

\subsubsection{Success probability under special favourable conditions}\label{SM_sec:special_cond}

Now, suppose at least one of the following three conditions hold: 
\begin{enumerate}
\item The Kraus operators are all mutually commuting, i.e. $[K_i, K_j] = 0$ for all $i, j = 0, 1, ..., R$.
\item $|\phi\rangle$ is stationary under the channel, i.e., $K_i | \phi \rangle \propto | \phi \rangle$ for all $i = 0, 1, ..., R$.
\item The circuit has a noise bias, and furthermore is bias preserving, such that we can choose some $|\phi\rangle$ to be immune to the dominant error (e.g. if the circuit only has phase flips, we can choose $|\phi \rangle = | 0 \rangle$).
\end{enumerate}

Under any of these conditions, we will find that
\begin{align*}
  &\frac{1}{T^2} \sum_{\mathbf{i}} \sum_{t = 1}^{T}  \sum_{q > t}^{T} \left(  \langle \psi | K_{i_q}^{\dag}K_{i_t} | \psi \rangle {\rm Tr} \rho_{\phi} \overline{K}_{i_q}^{\dag} \overline{K}_{i_t} + h.c. \right) \\
  &= \frac{T - 1}{T} \sum_{i, j} \langle \psi | K_i^{\dag} K_j | \psi \rangle \langle \phi | K_j^{\dag} K_i | \phi \rangle
\end{align*}

To make further progress, we note that not all trajectories containing Kraus operators where $i \neq 0$ or $j \neq 0$ are necessarily rejected by the protocol. Hence assuming that all such trajectories are rejected will give us a lower bound on the success probability, given by
\begin{align*}
  P_S^{(\log T)}
  &\geq \frac{1}{T} + \frac{T - 1}{T}  \langle \psi | K_0^{\dag} K_0 | \psi \rangle \langle \phi | K_0^{\dag} K_0 | \phi \rangle \\
  &\geq 1 - 4 \varepsilon + \frac{4 \varepsilon}{T} \\
  &\geq 1 - 4 \varepsilon
\end{align*}

\subsubsection{General fidelity scaling}

Having established that in the small parameter case, we can get $1 - P_s \sim O(\varepsilon)$ either under the conditions of the previous section, or when $T$ is not too large, we can now proceed to compute the fidelity associated with having $\log T$ control qubits.

In terms of $\tilde{\rho}^{(\log T)}, P_S^{\log T}$,
\begin{equation}
  \begin{aligned}
    (1 - F)_{\log T}
    &=  \frac{P_S^{\log T} - \langle U \psi | \tilde{\rho}^{(\log T)} | U \psi \rangle}{P_S^{\log T}},
  \end{aligned}
\end{equation}
where we can expand, 
\begin{align*}
  \langle U \psi | \tilde{\rho}^{(\log T)} | U \psi \rangle
  &= \frac{1}{T}(1 - F)_0
    + \frac{1}{T^2} \sum_{\mathbf{i}} \sum_{t = 1}^{T}  \sum_{q \neq t}  \left(  \langle \psi | U^{\dag} K_{i_t} | \psi \rangle \langle \psi | K_{i_q}^{\dag} U | \psi \rangle {\rm Tr} \rho_{\phi} \overline{K}_{i_q}^{\dag} \overline{K}_{i_t} \right)
\end{align*}

Since we are calculating quantities only to $O(\varepsilon)$, in the numerator we only need to keep compute terms for which either all the applied Kraus operators are $K_0$, or all the applied Kraus operators but one are $K_0$ (we can loosen this assumption when dealing with the bounded probability case).

First, we can combine the two terms in the numerator by factoring out the $\phi$ associated terms:
\begin{align*}
  &P_S^{\log T} - \langle U \psi | \tilde{\rho}^{(\log T)} | U \psi \rangle \\
  &= \frac{1}{T} (1 - F)_0
    + \frac{1}{T^2} \sum_{\mathbf{i}} \sum_{t = 1}^T \sum_{q \neq t} \left( \langle \psi | K_{i_q}^{\dag} K_{i_t} | \psi \rangle - \langle \psi | U^{\dag} K_{i_t} | \psi \rangle \langle \psi | K_{i_q}^{\dag} U | \psi \rangle \right) {\rm Tr} \rho_{\phi} \overline{K}_{i_q}^{\dag} \overline{K}_{i_t}
\end{align*}

Consider the terms in the brackets. For the terms in the sum where all Kraus operators applied are $K_0$, we have 
\begin{align*}
  &\langle \psi | K_0^{\dag} K_0 | \psi \rangle - \langle \psi | U^{\dag} K_0 | \psi \rangle \langle \psi | K_0^{\dag} U | \psi \rangle \\
  &= 1 - \varepsilon \langle \psi | U^{\dag} \xi | \psi \rangle - \varepsilon \langle \psi | \xi^{\dag} U | \psi \rangle - 1 + \varepsilon \langle \psi | U^{\dag} \xi | \psi \rangle + \varepsilon \langle \psi | \xi^{\dag} U | \psi \rangle + O(\varepsilon^2) \\
  &= O( \varepsilon^2 ),
\end{align*}
so these terms don't contribute.

For the terms where $i_q = 0$ and $i_t \neq 0$, we have 
\begin{align*}
  &\langle \psi | K_0^{\dag} K_{i_t} | \psi \rangle - \langle \psi | U^{\dag} K_{i_t} | \psi \rangle \langle \psi | K_0^{\dag} U | \psi \rangle \\
  &= \langle \psi | U^{\dag} K_{i_t} | \psi \rangle - \langle \psi | U^{\dag} K_{i_t} | \psi \rangle + O(\varepsilon^{3/2}) \\
  &= O(\varepsilon^{3/2}).
\end{align*}
Now, we can neglect the $O(\varepsilon^{3/2})$ term, since multiplying by the remaining $\phi$-associated factor $O(\varepsilon^{1/2})$, it will be $O(\varepsilon^2)$. A similar cancellation occurs for the terms where $i_q \neq 0$ and $i_t = 0$. We are left with 
\begin{align*}
P_S^{\log T} - \langle U \psi | \tilde{\rho}^{(\log T)} | U \psi \rangle 
  &= \frac{1}{T} (1 - F)_0 + O(\varepsilon^2)
\end{align*}

Finally, since $(1 - F)_0 \sim O( \varepsilon)$ and $P_S^{\log T} \sim 1 - O(\varepsilon)$,
\begin{equation}
  \begin{aligned}
    (1 - F)_{\log T}
    &=  \frac{P_S^{\log T} - \langle U \psi | \tilde{\rho}^{(\log T)} | U \psi \rangle}{P_S^{\log T}} \\
    &= \frac{\frac{1}{T}( 1 - F)_0}{1 - O(\varepsilon)} + O(\varepsilon^2) \\
    &= \frac{1}{T}(1 - F)_0 + O(\varepsilon^2),
  \end{aligned}
\end{equation}
which gives Eqn.~[3] of the main text.

\subsection{Unitary errors} \label{SM_sec:coherent_error}

This calculation comes with one significant caveat. The way we have written $K_0 = U - \varepsilon \xi$ looks extremely general, and one is tempted to conclude that this error suppression works for any such Kraus channel. However, we note that the behaviour for unitary error channels, where $K_0 = V$ is a pure unitary is exceptional and must be considered separately. To see why that is the case, suppose $V = U - \varepsilon \xi$ is unitary. Then 
\begin{align*}
  V V^{\dag}
  &= (U - \varepsilon \xi) (U^{\dag} - \varepsilon \xi^{\dag} ) \\
  &= 1 - \varepsilon (U \xi^{\dag} + \xi U^{\dag}) + O(\varepsilon^2).
\end{align*}

In other words, for $V$ to be unitary to first order we require $U \xi^{\dag} + \xi U^{\dag} = 0$. Now, the base fidelity of such a channel is, 
\begin{align*}
  (1 - F)_0
  &= 1 - \langle \psi | U^{\dag} V |\psi \rangle \langle \psi | V^{\dag} U | \psi \rangle \\
  &= \varepsilon (U \xi^{\dag} + \xi U^{\dag} ) + O( \varepsilon^2) \\
  &= O(\varepsilon^2).
\end{align*}
Since $(1 - F)_0 \sim O(\varepsilon^2)$, the calculation of the previous section tells us nothing about the error suppression of such unitary error channels.

In fact, coherent unitary errors cannot be suppressed by gate-based EF. Suppose $\mathcal{U}(\cdot) = V (\cdot) V^{\dag}$. It is not difficult to see that the outcome of gate-based EF for any number of ancillas is always $\rho_{\log T} = V | \psi \rangle \langle \psi | V^{\dag}$, hence $F_{\log T} = F_0$. We conclude that gate-based EF works to suppress stochastic errors only, leaving unitary errors alone. Fortunately, this still encompasses the vast majority of error channels commonly considered.

\section{Guaranteeing fidelity enhancement with a single control qubit}\label{app_sec:guarantee}

We claim that error filtration can enhance fidelity with a single qubit as long as the base fidelity $F_0$ of the apparatus for the input state $| \psi \rangle$ satisfies $F_1 > \frac{1}{2}$. This can be done by setting $| \phi \rangle = | \psi \rangle$. The final state $\rho_1$ with a single control qubit is
\begin{align*}
  \rho_1 &= \frac{1}{2} \rho_0 + \frac{1}{2} \rho_0^2
\end{align*}
where $\rho_0 = \mathcal{U}( | \psi \rangle \langle \psi | )$.

The expression for the fidelity is 
\begin{equation}
F_1 = \frac{\langle U \psi | \rho_1 | U \psi \rangle}{{\rm Tr} \rho_1} = \frac{F_0 + \langle U \psi | \rho_0^2 | U \psi \rangle}{1 + {\rm Tr} \rho_0^2}.
\end{equation}

The condition $F_1 > F_0$ then reduces to
\begin{equation}
\langle U \psi | \rho_0^2 | U \psi \rangle - F_0 {\rm Tr} \rho_0^2
\end{equation}

To evaluate this, let us write 
\begin{equation}
\rho_0 = F_0 | U \psi \rangle \langle U \psi | + ( 1 - F_0 ) \sigma,
\end{equation}
where $\sigma$ is a density matrix satisfying $\langle U \psi | \sigma | U \psi \rangle = 0$. The relevant quantities become
\begin{align}
 \langle U \psi | \rho_0^2 | U \psi \rangle
  &= F_0^2 + ( 1 - F_0)^2 \langle U \psi | \sigma^2 | U \psi \rangle \\
  {\rm Tr} \rho_0^2
  &= F_0 + (1 - F_0)^2 {\rm Tr} \sigma^2
\end{align}

Then, 
\begin{equation}
\begin{aligned}
  \langle U \psi | \rho_0^2 | U \psi \rangle - F_0 {\rm Tr} \rho_0^2
  &= (1 - F_0) \left[  F_0^2 + (1 - F_0)^2 \left( \langle U \psi | \sigma^2 | U \psi \rangle  - F_0 {\rm Tr} \sigma^2  \right) \right] \\
  &\geq (1 - F_0) F_0 ( 2 F_0 - 1) > 0,
\end{aligned}
\end{equation}
where we have used $ \langle U \psi | \sigma^2 | U \psi \rangle \geq 0,  {\rm Tr} \sigma^2  \leq 1$. The above condition is satisfied whenever $F_0 > 1/2$, with $F_0 = F_1$ for a coherent channel. Hence, a channel improvement is guaranteed as long as the initial fidelity is greater than $1/2$, representing a minimal condition for which gate-based EF yields some improvement.

\section{Error floor for bounded failure probability}

As noted above the expressions do not lend themselves to easy simplification of an arbitrary $|\phi\rangle$, and we are unable to make strong statements about the behaviour of the gate-based EF circuit when $T \gg 1/\varepsilon$ in the general case. However, in the case of channels which admit a $| \phi \rangle$ for which we can write $K_i | \phi \rangle \propto | \phi \rangle$, not only is the probability of failure bounded (following conditions 2. and 3. from the previous section), one can show that the state approaches a well-defined limit as $T \rightarrow \infty$. We will show this implies that the $O(\varepsilon^2)$ terms do not blow up as $T \rightarrow \infty$, and represent an error floor to the scheme.

\subsection{Relation to vacuum extension}\label{SM_sec:pseudovac}

The assumption we make for this section is closely related to the concept of a \emph{vacuum extension}, introduced and defined Sec.~(2c) of \cite{chiribella_quantum_2019}. The vacuum extension provides a consistent way to define a superposition of channels. As \cite{chiribella_quantum_2019} points out, the original error filtration scheme \cite{gisin_error_2005} is such an example of a superposition of channels. Hence, one might take the point of view of error filtration as the operationalization of the superposition of channels to suppress errors. In gate-based EF, as demonstrated in our main results, we have loosened this demand, and do not require our scheme to strictly carry out a superposition of channels in order to succeed.

We note that any vacuum extension will yield the favourable probability scaling in gate-based EF. Furthermore, there are various physical systems in which a vacuum extension arises naturally, most notably in optical systems, where the vacuum extension of a system tends to be the actual electromagnetic vacuum. For instance, one might encode qubits with polarized light $|0 (1) \rangle = | H (V) \rangle $. Since most apparatuses act trivially on the electromagnetic vacuum, i.e. no input state, one can then define the vacuum extension to a channel using the actual electromagnetic vacuum.

For our purposes of the calculations in this section, we will not make full use of a rigorous definition of the vacuum extension. Instead, it will be sufficient for us to define a state $| \phi \rangle$ such that, 
\begin{equation}
K_i | \phi \rangle = \sqrt{q_{\phi}^{(i)}} | \phi \rangle,
\end{equation}
where $q_{\phi}^{(i)}$ can be interpreted as the probability of the Kraus operator $K_i$ occurring given $| \phi \rangle$ as an input state to the apparatus. For convenience, We will refer to any state $| \phi \rangle$ with these properties relative to our channel of interest as a \emph{pseudo-vacuum state}.

\subsection{Mixed unitary error channels}

\subsubsection{Two unitaries}

We first consider the simplest possible model where $K_0 = \sqrt{1 - p} U$ and $K_1 = \sqrt{p} V$, where $U, V$ are unitary operators. We assume the existence of a psuedo-vacuum state $| \phi \rangle$ relative to this channel,
\begin{equation}
K_0 | \phi \rangle = \sqrt{1 - p} | \phi \rangle, \quad \quad K_1 | \phi \rangle = \sqrt{p} | \phi \rangle.
\end{equation}

Now suppose we are given an input state $| \psi \rangle$. To be as restrictive as possible, we can pick the minimum fidelity state $| \psi \rangle$ in the input space, but this will not matter for our calculation. We can express the action of $V$ on $| \psi \rangle$ as: 
\begin{equation}
V | \psi \rangle = e^{ i \nu} \cos \theta | U \psi \rangle + \sin \theta | U \psi^{\perp} \rangle.
\end{equation}
with $\nu$ chosen so that $\theta \in [0, \pi / 2]$.  Note that we have to keep this phase around because it affects the outcome of gate-based EF when $T > 1$. $| U \psi^{\perp} \rangle$ is simply the part of the state that is orthogonal to $| U \psi \rangle$.

The fidelity for this state is
\begin{equation}
F_0 = 1 - p + p \cos^2 \theta = 1 - p \sin^2 \theta.
\end{equation}

Now we want to evaluate $F_{\infty}$, which is the fidelity approached by gate-based EF as $T \rightarrow \infty$. As $T \rightarrow \infty$, the output density matrix turns out to approach a pure state:
\begin{align}
  | \psi_{\infty} \rangle
  &\propto (1 - p) U | \psi \rangle + p V | \psi \rangle \\
  &= (1 - p + p e^{i \nu} \cos \theta) | U \psi \rangle + p \sin \theta | U \psi^{\perp} \rangle.
\end{align}

To see why this is the case, consider the density matrix for arbitrary $T$. Consider a single error configuration, where for the some set of branches $I_0$ where $|I_0| = m$, the black box carries out $K_0$ and in the rest of the $T - m$ branches it carries out $K_1$. Prior to measuring the control register, this corresponds to the state
\begin{align}
  &\frac{1}{\sqrt{T}} \sum_{t \in I_0} | t \rangle K_0 | \psi \rangle K_0^{m - 1} K_1^{T - m} | \phi \rangle + \frac{1}{\sqrt{T}} \sum_{t \notin I_0} | t \rangle K_1 | \psi \rangle K_0^m K_1^{T - m - 1} | \phi \rangle \\
  &= \frac{1}{\sqrt{T}} (1 - p)^{m / 2} p^{(T - m)/2} \left(  \sum_{t \in I_0} | i \rangle U | \psi \rangle + \sum_{t \notin I_0 } | i \rangle V | \psi \rangle  \right)| \phi \rangle.
\end{align}

Tracing out the vacuum register and measuring the control register, we are left with 
\begin{equation}
\frac{1}{\sqrt{2 T}} (1 - p)^{m / 2} p^{(T - m)/2} \left( m U | \psi \rangle + (T - m) V | \psi \rangle \right)
\end{equation}

We see that since the action on the pseudo-vacuum state is trivial, this state only depends on the number of times $K_0, K_1$ are applied in this particular error configuration. Thus all error configurations with the same $m, T - m$ distribution of Kraus operators coherently combine in the final density matrix, giving a $T$ choose $m$ enhancement to this state, such that the final (unnormalized) density matrix is 
\begin{equation}
\rho_T = \sum_{m = 1}^{T} \frac{1}{2 T} {T \choose m} (1 - p)^{m} p^{T - m} \left( m U | \psi \rangle + (T - m) V | \psi \rangle \right)  \left( m U | \psi \rangle + (T - m) V | \psi \rangle \right)^{\dag}
\end{equation}

Now, we know from the binomial distribution that as $T \rightarrow \infty$ that when viewed as a function of $m$, ${T \choose m} (1 - p)^m p^{T - m}$ is sharply peaked around $m = (1 - p) T$. Thus for $T \rightarrow \infty$, 
\begin{equation}
\rho_{\infty} \simeq  \left( (1 - p) U | \psi \rangle + p V | \psi \rangle \right)  \left( (1 - p) U | \psi \rangle + p  V | \psi \rangle \right)^{\dag}
\end{equation}
which is the pure state $| \psi_{\infty} \rangle$. This provides a rigorous sense in which we are converting a mixture of states $U | \psi \rangle, V | \psi \rangle$ into a superposition. The trace and fidelity of this state are
\begin{align}
  {\rm Tr} \rho_{\infty} &= (1 - p + p e^{i \nu} \cos \theta)^2 + p^2 \sin^2 \theta, \\
  \langle U \psi | \rho_{\infty} | U \psi \rangle &= (1 - p + p e^{i \nu} \cos \theta)^2, \\
  F_{\infty} &= \frac{(1 - p + p e^{i \nu} \cos \theta)^2}{(1 - p + p e^{i \nu} \cos \theta)^2 + p^2 \sin^2 \theta}.
\end{align}

We are interested in two questions. First, when does this represent an improvement on the base fidelity -- in other words, when is $F_0 < F_{\infty}$? Second, in the small parameter case, what does this mean for the $O(\varepsilon^2)$ terms?

First, if $\nu$ is known, then we can simply write down $F_0 < F_{\infty}$ in terms of $p, \nu$ as
\begin{equation}
1 - p + p \cos^2 \theta < \frac{(1 - p + p e^{i \nu} \cos \theta)^2}{(1 - p + p e^{i \nu} \cos \theta)^2 + p^2 \sin^2 \theta}.
\end{equation}
Note that expression, unlike the single qubit control case, does not furnish a condition on $F_0$, since $F_0$ contains no information about $\nu$. It will be useful for us to pare down this requirement even more to a condition stated only in terms of $p$, since $p, 1 - p$ are easily determined.

It is instructive to consider the two extremal values $\nu = 0, \pi$. $\nu = 0$ corresponds to constructive interference in the final state and $\nu = \pi$ corresponds to destructive inteference. The $\nu = 0$ case showcases how phase coherence can amplify the effects of error filtration. However, we will make the pessimistic assumption that $\nu = \pi$ in order to obtain a minimal and single-variable success condition, since
\begin{equation}
F( | \psi_{\infty} \rangle ) > F( (1 - p - p \cos \theta) | U \psi \rangle + p \sin \theta | U \psi^{\perp} \rangle) \equiv F_{\infty}',
\end{equation}
where $F(\cdot)$ refers to the fidelity of the state $(\cdot)$. With this, the pessimistic success condition is
\begin{equation}
1 - p + p \cos^2 \theta = F_1 < F_{\infty}' = \frac{(1 - p - p \cos \theta)^2}{ (1 - p - p \cos \theta)^2 + p^2 \sin^2 \theta}.
\end{equation}

To simplify, we write $\cos \theta$ in terms of $F_1$ and solve for $p$. Subject to the constraints $0 \leq p \leq 1, 1 - p \leq F_1 < 1$, this yields 
\begin{equation}
p < \frac{1}{4}, \quad \mbox{or}
\end{equation}
\begin{equation}
\frac{1}{4} < p < \frac{1}{2} \mbox{ and } 1 - p < F < \frac{1}{4p}.
\end{equation}

The first condition is easy to interpret. The inequality $p < \frac{1}{4}$ asserts that the probability of $V$ is sufficiently small that the final state resembles $| U \psi \rangle$ despite possible destructive interference due to $\nu$. The second condition provides for the possibility that the interference can be quite possibly be quite large, given both a destructive phase and a larger probability of $V$. However, the upper bound on $F$ in that case serves to bound the size of $\cos \theta$ and thus the size of the destructive interference.

In terms of $F_1$, this establishes two regions of success, although $F_1$ alone cannot yield a sufficient condition over this entire range. As we will see shortly, the first condition generalizes more readily to more complicated channels.

Finally, let us consider where the small parameter regime fits in this picture. Let $2 p = \varepsilon \ll 1$ so that $K_0$ is of the form $U - \varepsilon A + O(\varepsilon^2)$. Then we can expand $F_{\infty}$ as
\begin{equation}
\begin{aligned}
  F_{\infty}
  &= \frac{1 - \varepsilon ( 1 + e^{i \nu} \cos \theta ) + \frac{1}{4} \varepsilon^2( 1 + e^{i \nu} \cos \theta )^2 }{1 - \varepsilon ( 1 + e^{i \nu} \cos \theta ) + \frac{1}{4} \varepsilon^2[( 1 + e^{i \nu} \cos \theta )^2 + \sin^2 \theta]} \\
  &= 1 - \frac{1}{4} \sin^2 \theta \varepsilon^2 + O( \varepsilon^2).
\end{aligned}
\end{equation}
This reaffirms the small parameter scaling we derived earlier, as $F_{T \rightarrow \infty} \rightarrow 1 - O(\varepsilon^2)$. Furthermore, the coefficient of $\varepsilon^2$ as $T \rightarrow \infty$ approaches a $T$ independent limit and does not blow up. Hence, we can interpret it as an error floor to gate-based EF under the assumptions in this section.

\subsubsection{Many unitaries, with $K_0$ perfect}

Now we consider the case of a mixed unitary error channel with $K_0 \propto U$ and multiple unitary errors possible, so that
\begin{equation}
K_0 = \sqrt{1 - p} U, \quad \quad K_{i \neq 0} = \sqrt{p \lambda_i} V_i, 
\end{equation}
where $\sum_i \lambda_i = 1$. We write the action of $V$ on $| \psi \rangle$ as 
\begin{equation}
V_i | \psi \rangle = e^{i \nu_i} \cos \theta_i | U \psi \rangle + \sin \theta_i | U \psi^{\perp}_i \rangle.
\end{equation}
Again, we choose $\nu$ such that $\theta \in [0, \pi / 2]$. The base fidelity is 
\begin{equation}
F_0 = 1 - p + p \sum_i \lambda_i \cos^2 \theta_i.
\end{equation}

Following the same argument as the two unitary case, we find 
\begin{equation}
  | \psi_{\infty} \rangle = (1 - p)  U |\psi \rangle + p \sum_i \lambda_i V_i | \psi \rangle
  = ( 1 - p + p \sum_i \lambda_i e^{i\nu_i} \cos \theta_i) | U \psi \rangle + p \sum_i \lambda_i \sin \theta_i | U \psi^{\perp}_i \rangle
\end{equation}

To derive a minimal success condition in the same spirit as before, we note that the fidelity is minimized when $\nu_i = - \pi$ for all $i$. This corresponds to constructive interference between all the $|U \psi_i^{\perp} \rangle$, which is greatest when they are all parallel, i.e. $|U \psi_i^{\perp} \rangle = |U \psi^{\perp} \rangle$ for all $i$. Let
\begin{equation}
| \psi_{\infty}' \rangle = ( 1 - p - p \sum_i \lambda_i \cos \theta_i) | U \psi \rangle + p \sum_i \lambda_i \sin \theta_i | U \psi^{\perp} \rangle.
\end{equation}
Then 
\begin{equation}
F( | \psi_{\infty} \rangle ) > F( | \psi_{\infty}' \rangle ) \equiv F'_{\infty},  
\end{equation}
and it will be sufficient for us to say $F_1 < F'_{\infty}$. We calculate $F'_{\infty}$:
\begin{equation}
F_{\infty}' = \frac{ ( 1 - p - p \sum_i \lambda_i \cos \theta_i)^2}{ ( 1 - p - p \sum_i \lambda_i \cos \theta_i)^2 +  p^2 \left( \sum_i \lambda_i \sin \theta_i \right)^2}.
\end{equation}
The $F_0 < F_{\infty}$ condition now reads: 
\begin{equation}
 1 - p + p \sum_i \lambda_i \cos^2 \theta_i < \frac{ ( 1 - p - p \sum_i \lambda_i \cos \theta_i)^2}{ ( 1 - p - p \sum_i \lambda_i \cos \theta_i)^2 +  p^2 \left( \sum_i \lambda_i \sin \theta_i \right)^2}.
\end{equation}

So far the above still contains $p, \theta_i, \lambda_i$. We'd like to reduce it to just $p$. To do so, we will wrestle it into the form we've already solved for the two unitary channel. To do so, we note the trivial bound
\begin{equation}
F_1 =  1 - p + p \sum_i \lambda_i \cos^2 \theta_i < F_1' = 1 - p + p \cos^2 \theta_{max}
\end{equation}
where $\cos \theta_{max}$ is the maximum possible value $\cos \theta_i$ can take. $F_{\infty}$ is similarly lower bounded by
\begin{equation}
F( ( 1 - p - p \sum_i \lambda_i \cos \theta_i) | U \psi \rangle + p \sum_i \lambda_i \sin \theta_i | U \psi^{\perp} \rangle ) > F( ( 1 - p - p \cos \theta_{max}) | U \psi \rangle + p \sin \theta_{max} | U \psi^{\perp} \rangle )
\end{equation}
where $\sin \theta_{max}$ is the minimum possible value of $\sin \theta_i$. Then our condition will be fulfilled if we have
\begin{equation}
 1 - p + p \cos^2 \theta_{max} < \frac{ ( 1 - p - p \cos \theta_{max})}{ ( 1 - p - p \cos \theta_{max})^2 +  p^2 \sin^2 \theta_{max}}
\end{equation}
which reduces to the two unitary case. Hence, applying the results of the previous section,
\begin{equation}
p < \frac{1}{4}
\end{equation}
is a minimal single parameter success condition.

\subsubsection{Many unitaries, with $K_0$ imperfect}

Our next generalization is to allow $K_0$'s action on $| \psi \rangle$ to have a perpendicular component $| U \psi^{\perp} \rangle$, such that 
\begin{equation}
K_0 | \psi \rangle = \sqrt{1 - p} \cos \theta_0 | U \psi \rangle + \sqrt{1 - p} \sin \theta_0 | U \psi^{\perp} \rangle.
\end{equation}
We assume that $\cos \theta_0 > 1 / 2$, otherwise one can hardly call the apparatus something that attempts to carry out the operation $U$. In particular, $\cos \theta_0 < 1 / 2$ would imply $F_1 < 1 / 2$, a situation which we will try to avoid entirely.

The action of $K_{i \neq 0}$ on $| \psi \rangle$ is defined as before. Again, we will deal with the two operator case first, for which
\begin{align}
F_0 &= (1 - p) \cos^2 \theta_0 + p \cos^2 \theta_1.
\end{align}

To derive a condition on $\mathcal{U}$, we will bound the limiting fidelity $F_{\infty}^{(K)}$ associated with $\mathcal{U}$ with the limiting fidelity $F_{\infty}^{(L)}$ associated with a different channel $\mathcal{L}$. Consider such a Kraus channel $\mathcal{L}$ with three operators $L_0, L_0', L_1$, whose action on $| \psi \rangle$ are defined by
\begin{align}
  L_0 | \psi \rangle &= \sqrt{(1 - p) \cos \theta_0} | U \psi \rangle \\
  L_0' | \psi \rangle &= \sqrt{(1 - p) \sin \theta_0} | U \psi^{\perp}_0 \rangle \\
  L_1 | \psi \rangle &= K_1 | \psi \rangle
\end{align}
where $L_0, L_0'$ are proportionate to unitaries, so that their action on the pseudo-vacuum extension is, by assumption,
\begin{align}
  L_0 | \phi \rangle &= \sqrt{(1 - p) \cos \theta_0}, \\
  L_0' | \phi \rangle &= \sqrt{(1 - p) \sin \theta_0}, \\
  L_1 | \phi \rangle &= K_1 | \phi \rangle.
\end{align}

First let's check if this channel is physical. We have
\begin{align}
  {\rm Tr} \left(  L_0 | \psi \rangle \langle \psi | L_0 \right)
  &= (1 - p) \cos^2 \theta_0 \\
  {\rm Tr} \left(  L_0' | \psi \rangle \langle \psi | L_0' \right)
  &= (1 - p) \sin^2 \theta_0 \\
  {\rm Tr} \left(  L_1 | \psi \rangle \langle \psi | L_1 \right)
  &= p
\end{align}
which add up to $1$. Hence, by defining the action of $L_i$ on the rest of the Hilbert space appropriately, we can construct a valid Kraus channel with these properties.

The fidelity of $\mathcal{L}$ on $| \psi \rangle$ is
\begin{equation}
F_0^{(L)} = (1 - p) \cos^2 \theta_0 + p \cos^2 \theta_1 = F_1
\end{equation}

The $T \rightarrow \infty$ limits for $\mathcal{U}, \mathcal{L}$ before measurement of the control register are given by the pure unnormalized states, 
\begin{equation}
\begin{aligned}
 | \psi_{\infty}^{(K)} \rangle
 &= \left[ (1 - p) \cos \theta_0 + p e^{i \nu_1} \cos \theta_1 \right] | U \psi \rangle + (1 - p) \sin \theta_0 | U \psi^{\perp}_0 \rangle + p \sin \theta_1 | U \psi_1^{\perp} \rangle \\
   | \psi_{\infty}^{(L)} \rangle
  &= \left[  (1 - p) \cos^2 \theta_0 + p  e^{i \nu_1} \cos \theta_1  \right] | U \psi \rangle + (1 - p) \sin^2 \theta_0 | U \psi^{\perp}_0 \rangle + p \sin \theta_1 | U \psi_1^{\perp} \rangle
\end{aligned}
\end{equation}

We observe that if $- \pi/2 < \nu < \pi/2$ implies that
\begin{equation}
F(| \psi_{\infty}^{(K)} \rangle) > F(| \psi_{\infty}^{(L)} \rangle)
\end{equation}
since $\cos \theta_0 - \cos^2 \theta_0 > \sin \theta_0 - \sin^2 \theta_0 > 0$, so $| \psi_{\infty}^{(L)} \rangle$ has a relatively larger component in $| U \psi \rangle$. If $\pi / 2 < \nu < 3\pi / 2$, then we just have to ensure $(1 - p) \cos \theta_0 > p \cos \theta_1$ for the same thing to be true, which can be ensured by $1 - p > p$.

Thus we only have to show that the condition holds for the channel specified by Kraus operators $L_i$. We have done this already, and re-using the minimal condition we found earlier with $( 1 - p) \mapsto (1 - p) \cos^2 \theta_0$, we have the condition 
\begin{equation}
(1 - p) \cos^2 \theta_0 > \frac{3}{4},
\end{equation}
which straightforwardly generalizes to as Kraus operators $K_{i \neq 0}$ as we want.

\subsection{Channels with non-unitary Kraus operators}

Finally, we deal with the most general case, where $K_i$ can be non-unitary and $K_0$ can be imperfect. In that case the probability of $K_i$ is state dependent. The key insight here is to realize that as $T \rightarrow \infty$, the Kraus operators have well-defined probability given by 
\begin{equation}
q_{\phi}^{(i)} \equiv \langle \phi | K_i^{\dag} K_i | \phi \rangle,
\end{equation}
because at each time step, we feed the apparatus $T - 1$ copies of $| \phi \rangle$ and only a single copy of $| \psi \rangle$. Note that $K_i$ is effectively unitary on $| \phi \rangle$.

To generalize our previous results, we simply have to replace $1 - p$ with $q_{\phi}^{(0)}$ and $p \lambda_i$ with $q_{\phi}^{(i)}$ in the expression for $F_{\infty}$. If $q_{\phi}^{(0)} = 1 - p$ as with the unitary case, then $q_{\phi}^{(0)}  > 3 / 4$ will suffice. Since $K_0 | \psi \rangle$ has the largest component of $| U \psi \rangle$, if $q_{\phi}^{(0)} > 1 - p$ can only increase $F_{\infty}$. Thus our condition becomes
\begin{equation}
q_{\phi}^{(0)} > q_{\psi}^{(0)} > \frac{3}{4 \cos^2 \theta_0}.
\end{equation}

\section{Isomorphism with error filtration for quantum communication}

In the event we have a psuedo-vacuum state $|\phi\rangle$ available to us (see Sec.~\ref{SM_sec:pseudovac}), our scheme is explicitly isomorphic to the original error filtration proposal in the ideal case \cite{gisin_error_2005, Vijayan2020robustwstate}. We will characterize this in the $T = 2$ case, with the more general $T$ following straight-forwardly. For simplicity, suppose our ideal unitary is the identity, as is the case for a quantum communication task. Let the state we want to transmit be 
\begin{equation}
| \psi \rangle = \alpha | 0 \rangle + \beta | 1 \rangle.
\end{equation}

The effect of multiplexing in error filtration is to encode this in a W-state \cite{Vijayan2020robustwstate}. For $T = 2$, this becomes 
\begin{equation}
  \alpha \hat{a}_0^{\dagger} + \beta \hat{a}_1^{\dagger}
  \mapsto \frac{1}{\sqrt{2}} \alpha \left(  \hat{a}_0^{\dagger} +  \hat{b}_0^{\dagger} \right) + \frac{1}{\sqrt{2}} \beta \left(  \hat{a}_1^{\dagger} +  \hat{b}_1^{\dagger} \right),
\end{equation}
where $\hat{a}_i, \hat{b}_i$ refer to some kind of encoding in a photonic mode.

Explicitly writing this out in the fock basis, we have 
\begin{equation}
  \frac{1}{\sqrt{2}} \alpha | 1, 0, 0, 0 \rangle
  +  \frac{1}{\sqrt{2}} \beta | 0, 1, 0, 0 \rangle
  +  \frac{1}{\sqrt{2}} \alpha | 0, 0, 1, 0 \rangle
  +  \frac{1}{\sqrt{2}} \beta | 0, 0, 0, 1 \rangle
\end{equation}

  In gate-based EF, we pair the state $ | \psi \rangle $ with $| + \rangle$ and attach $| \phi \rangle$ states, which looks like 
\begin{align}
  & \frac{1}{\sqrt{2}} | 0 \rangle \left( \alpha | 0 \rangle + \beta | 1 \rangle \right) | \phi \rangle + \frac{1}{\sqrt{2}} | 1 \rangle \left( \alpha | 0 \rangle + \beta | 1 \rangle \right) | \phi \rangle \\
  & \mapsto
    \frac{1}{\sqrt{2}} \alpha | 00 \rangle  | \phi \rangle
  +  \frac{1}{\sqrt{2}} \alpha | 10 \rangle  | \phi \rangle
  +  \frac{1}{\sqrt{2}} \beta | 01 \rangle  | \phi \rangle
  +  \frac{1}{\sqrt{2}} \beta | 11 \rangle  | \phi \rangle
\end{align}

Written in this way one can clearly see the one-to-one correspondence of kets in either scheme. Due to the principle of superposition, we can deal with each ket individually. If the error operators only act trivially on the state $| \phi \rangle$ (in gate-based EF) and the vacuum (in EF for quantum communication), then we can carefully map the effective error on each ket in gate-based EF to an error on each ket in error filtration. 

More formally it was observed in Eqn.~(3.1) of \cite{chiribella_quantum_2019}, there exists a unitary operation that connects a spatially separated quantum superposition of paths into a quantum superposition of paths over an additional register:
\begin{align}
  U ( | 0 \rangle \otimes | \psi \rangle  ) &= | \psi \rangle \otimes | \Omega \rangle \\
  U (| 1 \rangle  \otimes | \psi \rangle  ) &= | \Omega \rangle \otimes | \psi \rangle.
\end{align}
By appending the state $| \phi \rangle$, one can embed this in an isomorphism $U'$ that maps gate-based EF states to error filtration states.

\section{Ancilla errors}

\subsection{Effect of ancilla errors}

Finally, we work out the effect of ancilla errors. We will consider $X$ and $Z$ type errors, or bit flips and phase flips, on the control register and their effects on the fidelity and success probability of gate-based EF.

\begin{figure}[htbp]
\centerline{\includegraphics[width=\linewidth]{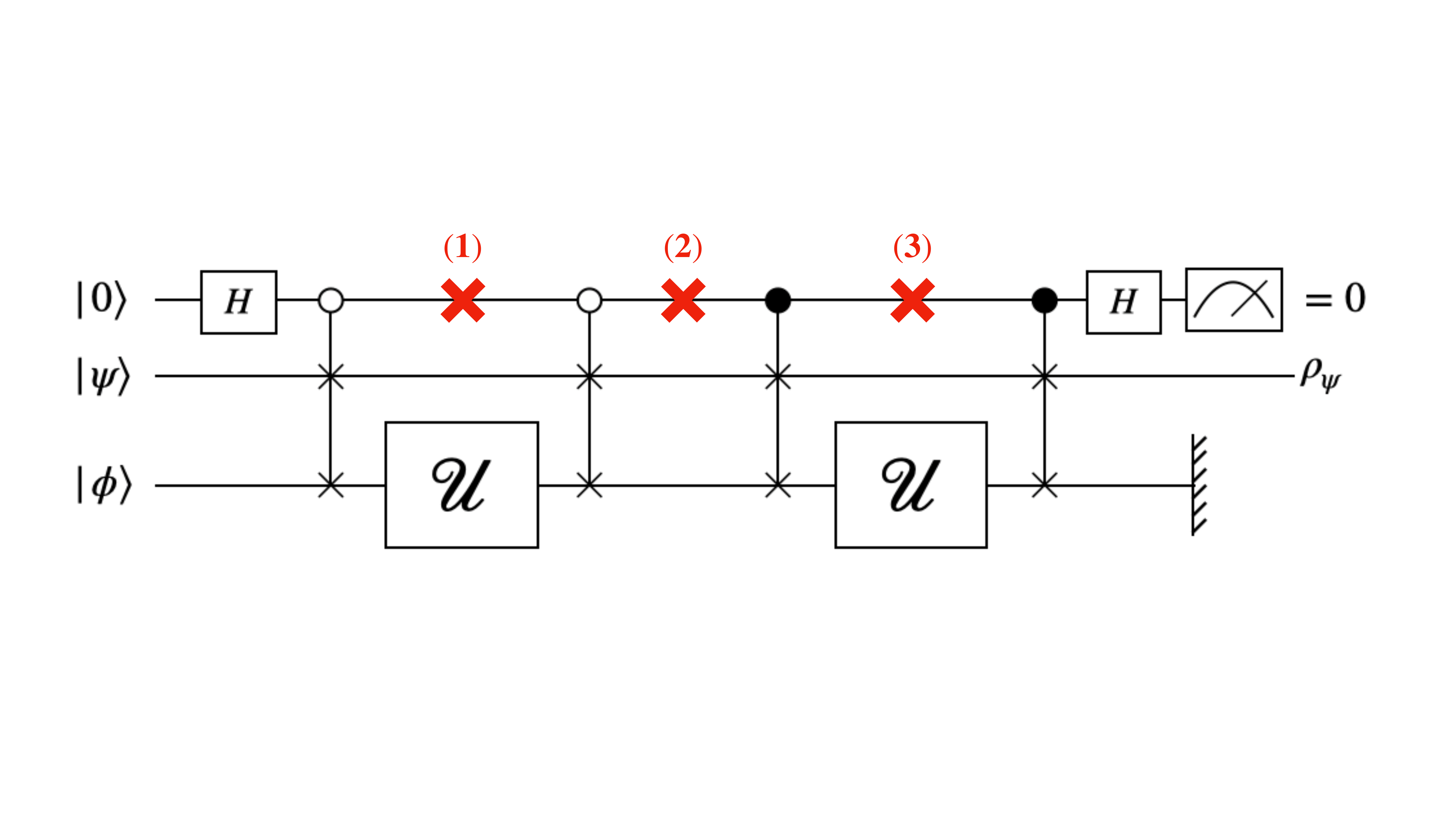}}
\caption[]{Gate-based EF with $T = 2$, with potential control fault locations highlighted in red.}
\label{fig:T2faults}
\end{figure}

\subsubsection{Bit flip errors}

A bit flip error on the ancilla can potentially ruin the whole query. To see why this happens, we consider the $T = 2$ case, for which the circuit is depicted in Fig.~\ref{fig:T2faults}. We consider the $3$ potential fault locations highlighted in red. For simplicity, suppose the bit flip errors occur in a run where the black box happens to apply the perfect unitary $U$ each time. We will show that a bit flip in any of these indicated fault locations in general leads to the failure of the query.

\begin{enumerate}
\item A bit flip in location $(1)$ leads to the state (before measurement and tracing out the active register) 
\begin{equation}
| + \rangle \otimes U | \phi \rangle \otimes U | \psi \rangle
\end{equation}
such that the final state is $U | \phi \rangle$, which in general is unrelated to the state we want. Since the control register is in the $| + \rangle$ state, an error in this location is maximally bad, as the final measurement will post-select $U | \phi \rangle$ with unit probability.

\item A bit flip in location $(2)$ leads to a double query of each of the participating states: 
\begin{equation}
\frac{1}{\sqrt{2}} |0 \rangle | \psi \rangle U^2 | \phi \rangle + \frac{1}{\sqrt{2}} | 1 \rangle U^2 |\psi \rangle | \phi \rangle,
\end{equation}
which does not contain our desired state $U | \psi \rangle$ at all.
\item A bit flip in location $(3)$ leads to the same final state as $(1)$.
\end{enumerate}

To estimate the effect of such errors on the final scheme, suppose the black box takes a time $\tau_U$ to run, and the error rate of each control ancilla is $\varepsilon_{bf}$. We assume that $\tau_U$ is much longer than the idle time of the control ancilla between the controlled-SWAPs on different branches, as well before and after the first and final queries.

For larger values of $T$, we can assume any such error of the type described by fault locations $(1), (3)$ have a $0$ contribution to the fidelity, and do not affect the post-selection probability.
To first order in $\varepsilon_C$ we expect the final infidelity to account for this as
\begin{equation}
(1 - F)_{\log T} \simeq \frac{1}{T} (1 - F)_0 + \tau_U \varepsilon_{bf} T \log T.
\end{equation}

Due to the disastrous potential of bit flip errors, we emphasize that gate-based EF should only be applied when the error hierarchy in the main text is respected with regard to bit flip errors on the control register. The above expression allows us to concretely estimate the error hierarchy required to handle bit flip errors. For gate-based EF to yield an overall error suppression, we require 
\begin{equation}
\frac{1}{T} (1 - F)_0 + \tau_U \varepsilon_{bf} T \log T < (1 - F)_0
\end{equation}

We can expressing the error hierarchy as a ratio between $\tau_U \varepsilon_{bf}$ and $(1 - F)_0$, we have
\begin{equation}
\frac{\tau_U \varepsilon_{bf}}{(1 - F)_0} < \frac{T - 1}{T^2 \log T} \simeq \frac{1}{T \log T}.
\end{equation}

This tells us that minimally, for $T = 2$, we need the error rate on ancilla qubits to be less than $1/2$ the base infidelity of the apparatus. Furthermore, this indicates that for a set-up with ancilla errors, there is an optimal $T$ beyond which gate-based EF no longer yields any advantage. This optimal $T$ is depicted for the case of QRAM in [FIG].

\subsubsection{Phase flip errors}

Since phase flip errors on the control register commute with any controlled-SWAP operation, a phase flip error on fault location $(1), (2)$ or $(3)$ will have the same effect. Let $\varepsilon_{pf}$ be the phase flip error rate on each control qubit. The presence of a phase flip error anywhere can be commuted to the end of the circuit, where the Hadamard transforms it into a bit flip error, upon which such an error causes the output of the circuit to be rejected.

This means that to first order in $\varepsilon, \varepsilon_{pf}$, phase flips \emph{do not affect the fidelity at all}. Since phase flip errors are detected, we expect there to be a decrease in success probability, to first order given by 
\begin{equation}
\Delta P_{pf} \simeq \tau_U \varepsilon_{pf} T \log T 
\end{equation}

In conclusion, the effect of phase flip and bit flip errors are well-characterized. Knowing the quantities $\tau_U, \varepsilon_{bf}, \varepsilon_{pf}, F_0$ then allows one to see whether the error hierarchy is respected and make a decision as to whether to carry out gate-based EF, as well as to what extent $T$.

\subsubsection{Memory errors}

A final category of ancilla error to consider are memory errors. These are most likely to happen to the memory register while the active register is querying the apparatus. Any error during this time is deleterious to the final query, as it will result in an error on $| \psi \rangle$ or $ U | \psi \rangle$ in the $T - 1$ inactive branches of the query. Such errors contribute in the same way to the fidelity and error hierarchy as bit flips.

\subsection{Numerical expansion of error suppressed QRAM with ancilla errors}

\begin{figure}[htbp]
\centerline{\includegraphics[width=0.6\linewidth]{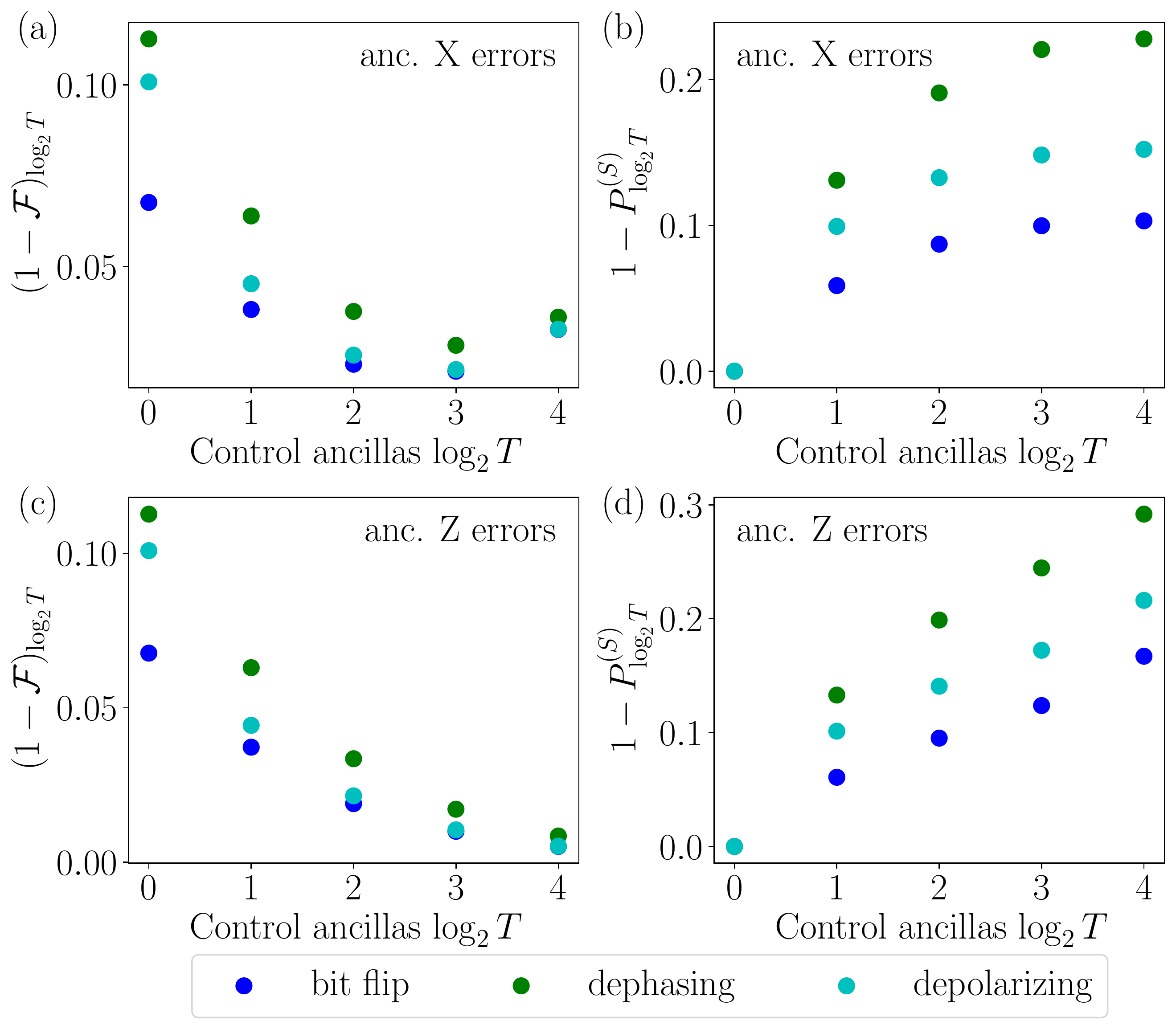}}
\caption[]{Gate-based EF applied to QRAM with ancilla errors, for QRAM depth $\log N = 2$. (a, b) Numerically estimated infidelity and probability of failure against number of control ancillas subject to $X$ errors at a rate $\varepsilon' = 0.001$. QRAM errors occur at rate $0.01$ per qubit per time step. (c, d) Same as (a, b) but for control ancilla subject to $Z$ errors.}
\label{fig:anc_err}
\end{figure}

Unfortunately, to perform a full simulation of both the gate-based EF applied to QRAM with errors in both ancilla and QRAM qubits is extremely computationally taxing. Instead, to observe the qualitative performance of gate-based EF under ancilla errors, it will suffice to perform a first order expansion of the infidelity and failure probabilities. 

To do so, we assume the infidelity and failure probabilities have the form \begin{align}
 (1 - F)_{\log_2 T} &\simeq (1 - N_{loc} \varepsilon') (1 - F)_{\log_2 T}^{(0)} +   \varepsilon' \sum_{\eta \in \mbox{err. loc.}} (1 - F)_{\log_2 T}^{(1), \eta}, \\
 P_{\log_2 T, {\rm fail}} &\simeq (1 - N_{loc} \varepsilon') P_{\log_2 T, {\rm fail}}^{(0)} +   \varepsilon' \sum_{\eta \in \mbox{err. loc.}}P_{\log_2 T, {\rm fail}}^{(1), \eta},
\end{align}
where $N_{loc}$ is the number of ancilla error locations in the circuit, $\varepsilon'$ is the probability of error per ancilla error location. The superscript $(0)$ indicates the bare infidelity and probability of failure without ancilla errors, as calculated in the main text. The superscript $(1), \eta$ requires a little more explanation. $\eta$ indexes ancilla error locations in the circuit, and the superscript $(1), \eta$ refers to the infidelity/probability of failure given an error in location $\eta$ but nowhere else. As such, the above expressions account for the first order expansion in ancilla errors, by discarding the possibility of errors occurring in both ancilla and QRAM at the same time. $(1 - F)_{\log_2 T}^{(1), \eta}, P_{\rm fail}^{(1), \eta}$ are then calculate numerically.

In this manner, expanding with $X, Z$ errors in the ancillas yields the plots in Fig.~\ref{fig:anc_err}. Note the qualitative features of these plots -- (1) Ancilla $X$ errors eventually overcome the error suppression from gate-based EF, and we have an optimal $T$ for which the overall infidelity is lowest. (2) Probability of failure still appears to plateau with ancilla $X$ errors. (3) Ancilla $Z$ errors do not to first order affect the infidelity scaling. However, this comes at the cost of increased failure probabilities.

\begin{figure}[htbp]
\centerline{\includegraphics[width=0.9\linewidth]{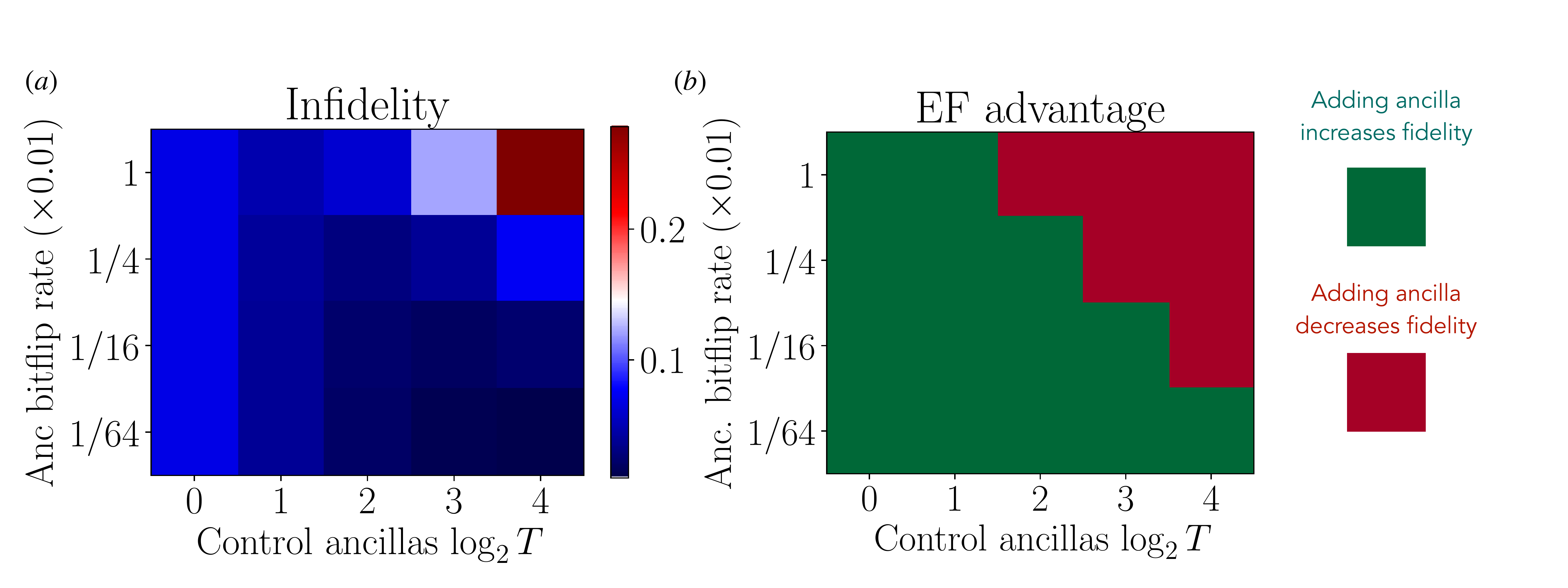}}
\caption[]{(a) Infidelity for a depth $2$ QRAM with gate-based EF against number of control ancillas (horizontal axis) and ancilla bitflip rate (vertical axis). (b) Plot of ``EF advantage" for the same data. The entry is green if adding the ancilla increases the fidelity for the given error rate, and red otherwise.}
\label{fig:EF_advantage}
\end{figure}

To illustrate the idea that there is an optimal $T$ for a given cSWAP and QRAM error rate, we plot infidelity against both ancilla bit flip rate (since only $X$ errors affect infidelity) and the number of control ancillas in Fig.~\ref{fig:EF_advantage}a. In Fig.~\ref{fig:EF_advantage}b, we make a qualitative version plot of the ``EF advantage", where a data point is depicted as green if adding one control ancilla improves the overall fidelity for a given ancilla bit flip rate. We observe that for each error rate, there is an optimal working point for $T$, given by the right-most green entry.

Using the perturbative expression, we can gain some insight into two questions. Given ancilla bit flip errors: (1) When does gate-based EF give an advantage with a single control qubit? (2) What is the maximum $T$ for which we still expect gate-based EF to yield better fidelities than the original black box operation?

For (1), we require that $(1 - F)_2 >  (1 - F)_0$. In this case, $N_{\rm loc} \sim O(1)$, and we write this as a constant $C = N_{\rm loc}$, and $(1 - F)^{(0)}_2 \simeq (1 - F)_0/2$. Conservatively, we can bound $ (1 - F)_{\log_2 T}^{(1), \eta} < 1$. Substituting these expressions into Eq.~(S81), we require \begin{equation}
    (1 - F)_0 > (1 - C \varepsilon') \frac{1}{2}(1 - F)_0 + C \varepsilon'.
\end{equation} 
Solving for $\varepsilon'$, we find that we require an ancilla bitflip rate \begin{equation}\label{app_eq:1C_cond}
    \varepsilon' < \frac{1}{2 C} (1 - F)_0 + O((1 - F)_0)^2 ,
\end{equation}
which demonstrates that we only need $\varepsilon'$ to be less than the infidelity of the black box up to a constant factor.

For (2), we require that $(1 - F)_{\log_2 T} >  (1 - F)_0$. In this case, the only difference from the above is that we have $N_{\rm loc} \sim O(T \log_2 T)$. Performing the same calculation, we find that we require 
\begin{equation}
    \varepsilon' < \frac{1}{O(T \log_2 T)} (1 - F)_0 + O ((1 - F)_0)^2  + O \left( \frac{1}{T} \right).
\end{equation}
Note that the above gives the condition for which having control ancillas still gives an improvement on the original black box fidelity. However, we only need to satisfy Eq.~(\ref{app_eq:1C_cond}) for gate-based EF to give some advantage.

\subsection{Mitigating bit flip errors with a flag qubit}

\begin{figure}[htbp]
\centerline{\includegraphics[width=0.75\linewidth]{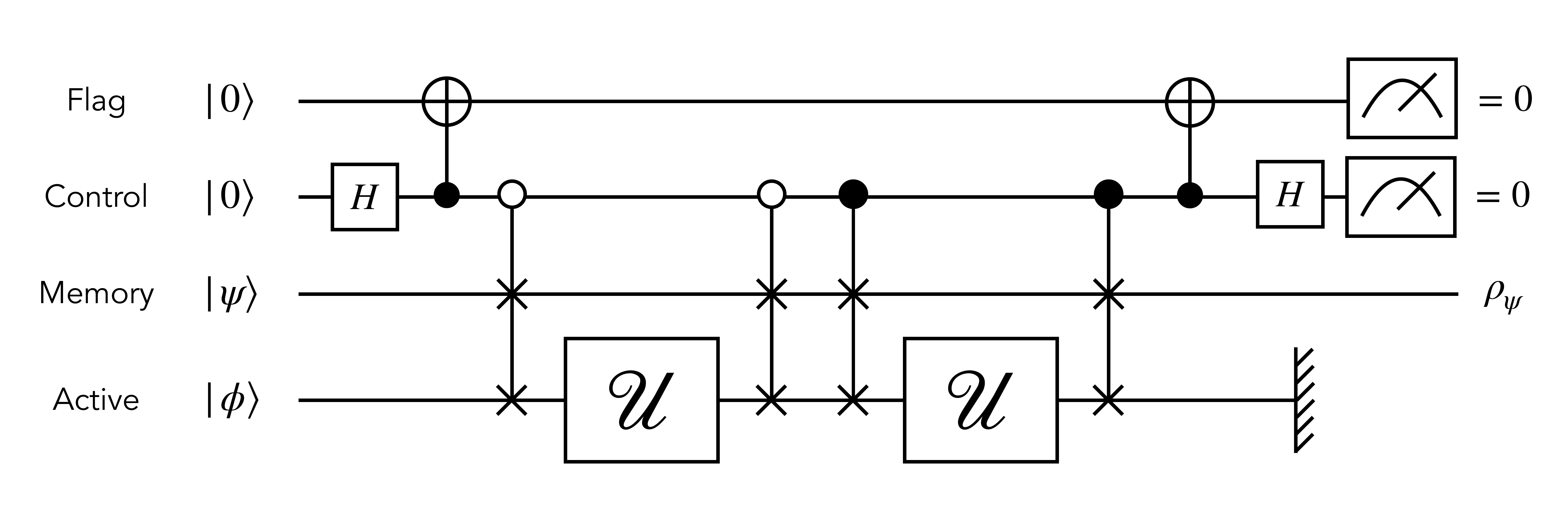}}
\caption{Gate-based EF with a single control qubit $(T = 2)$ and a flag qubit to detect bit flip errors in the ancilla.}
\label{fig:flag_qb}
\end{figure}

In the preceding discussion, we showed that bit flip errors on the control qubit can be problematic for gate-based EF. Fortunately, the ``control" part of the cSWAP acting on the ancilla commutes with the $Z$ operator. Hence, we can use a single flag qubit to detect bit flip errors in the ancilla \cite{flag_qb_ref_1}. In Fig.~\ref{fig:flag_qb}, the first hadamard and cNOT encode the flag and control in the $ZZ = +1$ subspace. The final cNOT allows us to measure the parity of the flag and control qubit. If the parity is odd, the flag measurement will yield $|1\rangle$, and we will know that a bit flip error must have occurred on either the flag or control qubit. In that case, we can discard the outcome and restart the procedure. This reduces the impact of bit flip errors on the fidelity quadratically (since two bit flip errors must occur for the bit flips to be undetectable), at the cost of a decreased probability of success. The flag qubit assisted scheme above generalizes easily to $\log T$ control qubits, by similarly appending a flag to each control qubit. This provides a way to mitigate the effect of bit flip errors on the ancilla with relatively low overhead.

\section{Optical implementations of the control register}

In this section, we point out several optical schemes for implementing the cSWAP operation crucial for gate-based EF. First, we note that the control register essentially acts as an addressing register in either a QRAM or Quantum Read-Only Memory (QROM) \cite{hann_resilience_2021}, with the memory register acting as the data to be queried, hence proposals (such as provided in the original BB-QRAM proposal \cite{giovannetti_quantum_2008}) for such quantum memories may be adapted to carry out the cSWAPs required for gate-based EF. Furthermore, several proposals for optical implementation of the cSWAP operations required for QRAM have been put forth since \cite{giovannetti_quantum_2008}, which show that it is possible with multiple different encodings. For instance, \cite{chen_scalable_2021} proposes an implementation using frequency-encoded optical qubits, and \cite{Hong_PRA_2012_robust_QRAM} proposes an implementation using fock state encoded qubits.

Finally, for concreteness, we highlight here an additional scheme by which one can carry out an optical cSWAP. The cSWAP operation can be decomposed into a fifty-fifty beam-splitter operation on the active and memory register, followed by a controlled-Z between the control and active register, followed by the inverse beam-splitter operation on the active and memory register (see Fig.~13b of \cite{chapman_cSWAP}). To achieve such an operation, we use optical polarization qubits for the active and memory registers, where the $|0\rangle, |1 \rangle$ logical states are encoded in the horizontal and vertical polarization states $|H\rangle, |V\rangle$. The beam-splitter operations can then be carried out using linear optics. Finally, to get a controlled-Z operation, we can use the first half of the Duan-Kimble scheme \cite{DuanKimble} by encoding the control qubit in a three-level atom, which will allow us to carry out a controlled-Z between the control qubit and the polarization-encoded memory register.

\break

\bigskip

\end{document}